\documentclass[12pt, a4paper, reprint, prl, aps, nofootinbib, nobibnotes, floatfix, showkeys, superscriptaddress, colorlinks]{revtex4-1}

\usepackage[utf8]{inputenc}
\usepackage{amsmath,amsthm,amssymb,amsfonts}
\usepackage{mathtools}
\usepackage{graphicx}
\graphicspath{{./figures/}}
\usepackage{booktabs}
\usepackage{xcolor}
\usepackage{rotating}
\usepackage{orcidlink}
\usepackage{adjustbox}
\usepackage{array}
\usepackage[mathlines]{lineno}
\usepackage{appendix}
\usepackage{aligned-overset}

\AtBeginDocument{
  \heavyrulewidth=.08em
  \lightrulewidth=.05em
  \cmidrulewidth=.03em
  \belowrulesep=.65ex
  \belowbottomsep=0pt
  \aboverulesep=.4ex
  \abovetopsep=0pt
  \cmidrulesep=\doublerulesep
  \cmidrulekern=.5em
  \defaultaddspace=.5em
}

\newcommand{\changed}[1]{#1}

\usepackage{hyperref}

\let\originalleft\left
\let\originalright\right
\renewcommand{\left}{\mathopen{}\mathclose\bgroup\originalleft}
\renewcommand{\right}{\aftergroup\egroup\originalright}
\newcommand{\abs}[1]{\left\lvert#1\right\rvert}

\newcommand{\norm}[1]{\left\|#1\right\|}

\newcommand{\re}[1]{\text{Re}\left[#1\right]} 
\newcommand{\im}[1]{\text{Im}\left[#1\right]} 
 
\newcommand{\var}[1]{\text{Var}\left[#1\right]}

\newcommand{\bmatrixByJames}[1]{\left[\;\begin{matrix}#1\end{matrix}\;\right]}

\newcommand{\intg}[3]{\int_{#1}^{#2} \text{d}#3\;}
\newcommand{\intginf}[1]{\intg{-\infty}{\infty}{#1}}

\newcommand{\h}[1]{\hat{#1}}

\newcommand{\R}{\mathbb{R}}

\newcommand{\Om}{\Omega}

\newcommand{\Si}{\Sigma}

\newcommand{\T}{\text{T}}

\newcommand{\barBetter}[1]{\mkern 1.5mu\overline{\mkern-1.5mu#1\mkern-1.5mu}\mkern 1.5mu}

\usepackage{outlines}

\begin{document}
\title{Achieving the fundamental quantum limit of linear waveform estimation}

\author{James~W.~Gardner\,\orcidlink{0000-0002-8592-1452}}
\email{james.gardner@anu.edu.au}
\affiliation{OzGrav-ANU, Centre for Gravitational Astrophysics, Research Schools of Physics, and of Astronomy and Astrophysics, The Australian National University, Canberra ACT 2601, Australia}
\affiliation{Walter Burke Institute for Theoretical Physics, California Institute of Technology, Pasadena, California 91125, USA} 
\author{Tuvia Gefen\,\orcidlink{0000-0002-3235-4917}}
\email{tgefen@caltech.edu}
\affiliation{Institute for Quantum Information and Matter, California Institute of Technology, Pasadena, CA 91125, USA}
\author{Simon A. Haine\,\orcidlink{0000-0003-1534-1492}}
\author{Joseph J. Hope\,\orcidlink{0000-0002-5260-1380}}
\affiliation{Department of Quantum Science and Technology and Department of Fundamental and Theoretical Physics, Research School of Physics, The Australian National University, Canberra ACT 0200, Australia}
\author{Yanbei Chen\,\orcidlink{0000-0002-9730-9463}\,}
\affiliation{Walter Burke Institute for Theoretical Physics, California Institute of Technology, Pasadena, California 91125, USA}

\date{\today}
\begin{abstract}
    Sensing a classical signal using a linear quantum device is a pervasive application of quantum-enhanced measurement. The fundamental precision limits of linear waveform estimation, however, are not fully understood. In certain cases, there is an unexplained gap between the known waveform-estimation Quantum Cram\'er-Rao Bound and the optimal sensitivity from quadrature measurement of the outgoing mode from the device. We resolve this gap by establishing the fundamental precision limit, the waveform-estimation Holevo Cram\'er-Rao Bound, and how to achieve it using a nonstationary measurement. We apply our results to detuned gravitational-wave interferometry to accelerate the search for post-merger remnants from binary neutron-star mergers. If we have an unequal weighting between estimating the signal's power and phase, then we propose how to further improve the signal-to-noise ratio by a factor of $\sqrt2$ using this nonstationary measurement.
\end{abstract}
\maketitle
\allowdisplaybreaks

In our efforts to probe fundamental physics, we invariably encounter the quantum limit: the irrevocable statistical nature of our reality~\cite{braginsky1995quantum,clerk2010introduction,danilishinQuantumMeasurementTheory2012}. This fundamental uncertainty of our measurement devices limits the precision at which we can sense classical signals.

\begin{figure}[ht]
    \centering
    \includegraphics[width=\columnwidth]{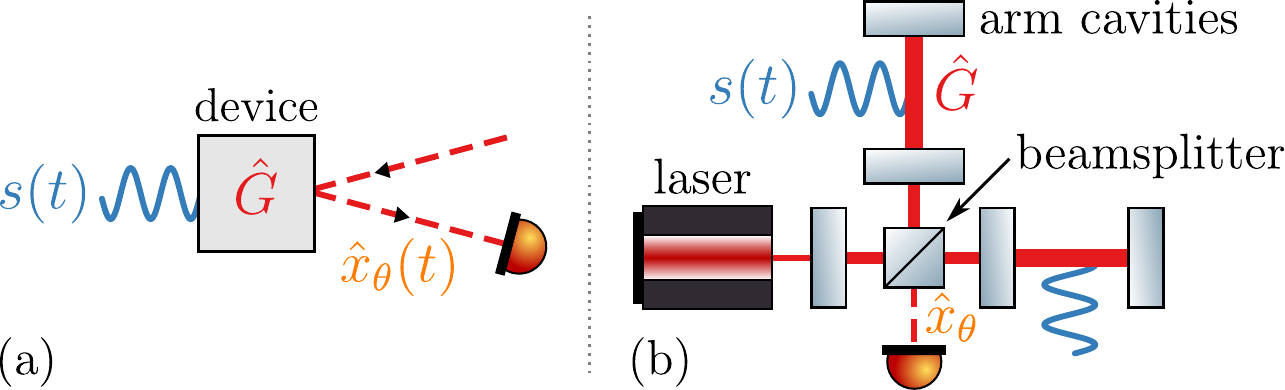}
    \caption{(a) A linear quantum device coupled to a classical signal. (b) For example, a gravitational-wave interferometer.}
    \label{fig:1}
\end{figure}

We consider the general problem of estimating a real classical signal $s(t)$ for all times $t$ using a linear quantum device as shown in Fig.~\hyperref[fig:1]{1a}~\cite{Tsang+2011}. The device evolves linearly according to a Hamiltonian $\h H(t) = \h H_0 + \h H_\text{int}(t)$ with the interaction $\h H_\text{int}(t) = -s(t)\h G$~\cite{Miao+2017}. Observables that do not commute with the generator $\h G$ (an internal degree-of-freedom) respond linearly to $s(t)$. We assume that the device is in a stationary state of $\h H_0$ which is time-invariant. In the input/output formalism~\cite{gardiner1985input}, information about the signal leaks out of the device into the environment imprinted on an {outgoing mode of a bosonic field.} By measuring this mode, we obtain a classical estimate of the classical signal mediated by the quantum device. \changed{The outgoing bosonic mode at each position and time is a harmonic oscillator} with canonical quadratures $\h x$ and $\h p$ which obey $[\h x, \h p]=i$ (let $\hbar=1$ henceforth). Let $\h x_\theta:=\cos(\theta)\h x + \sin(\theta)\h p$ for real $\theta$ such that the outgoing mode is~\cite{kubo1966fluctuation,buonanno2002signal} 
\begin{equation}\label{eq:IOrelation-TD}
    \h x_\theta(t) = \h x_\theta^{(0)}(t) + \intginf{t'} \chi_{x_\theta G}(t-t')s(t')
\end{equation}
\changed{where the superscript $(0)$ denotes the free evolution under $\h H_0$} and the susceptibility is $\chi_{x_\theta G}(t-t'):=i[\h x_\theta^{(0)}(t),\h G^{(0)}(t')]\Theta(t-t')$ with $\Theta$ the Heaviside function. In the Fourier domain, Eq.~\ref{eq:IOrelation-TD} becomes
\begin{equation}\label{eq:IOrelation-FD}
    \h x_\theta(\Om) = \h x_\theta^{(0)}(\Om) + \chi_{x_\theta G}(\Om)\tilde{s}(\Om)
\end{equation}
where the mode at each frequency $\Om$ is displaced by the signal's complex Fourier component $\tilde{s}(\Om):=\intginf{t} e^{i\Om t} s(t)$. Since $s(t)$, $\h x_\theta(t)$, and $\chi_{x_\theta G}(t)$ are real, their Fourier components obey $\tilde{s}(-\Om) = \tilde{s}^\dag(\Om)$ etc.\ such that it suffices to consider only positive frequencies.

Quantum metrology extends the classical theory of estimating parameters from a probability distribution. In particular, the Quantum Cram\'er-Rao Bound (QCRB)~\cite{braunstein1994statistical,WisemanMilburn2009book}, sets a fundamental precision limit: a lower bound on the variance of unbiased estimation of parameters encoded in a quantum state. This limit only depends on the state itself and not on the measurement scheme. In the real single-parameter case, the QCRB can always be saturated by the optimal measurement if the sample size is large~\cite{braunstein1994statistical}. Additionally, if the parameter appears as the shift in the mean of a Gaussian state, then the QCRB can be saturated for any sample size. In the multi-parameter case, however, reaching the QCRB is not guaranteed. In general, the QCRB can be saturated if and only if the symmetric logarithmic derivatives with respect to the parameters weakly commute~\cite{Holevo2011book}. For real parameters $s_j$ encoded as the shift in the mean of a Gaussian state by the unitary transformation $\exp(-i \sum_j s_j \h G_j)$, this condition is equivalent to the generators $\h G_j$ weakly commuting, i.e.\ $\langle[\h G_j, \h G_k]\rangle=0$, $\forall j,k$. 

Here, we want to simultaneously estimate the continuum of independent complex parameters $\tilde{s}(\Om)$ in Eq.~\ref{eq:IOrelation-FD}, one for each positive frequency $\Om$.

We assume that the linear device is at the quantum limit such that it is in a pure Gaussian state. The waveform-estimation QCRB $S_Q^\text{wave}$~\cite{Tsang+2011} sets a lower bound on the error $S_{ss}$ of unbiased estimation of $\tilde{s}(\Om)$ of

\begin{equation}\label{eq:waveform QCRB}
    \frac{S_{x_\theta x_\theta}(\Om)}{\abs{\chi_{x_\theta G}(\Om)}^2} =: S_{ss}(\Om) \geq S_Q^\text{wave}(\Om) := \frac{1}{S_{GG}(\Om)}
\end{equation}
where the power spectral density $S_{z_1z_2}$ for \textit{stationary} random processes $\h z_1$ and $\h z_2$ is defined as 
\begin{equation}
    2\pi\delta(\Omega-\Omega')S_{z_1z_2}(\Omega)=
    \langle\{\h z_1(\Om),\h z_2^\dag(\Om')\}\rangle.
\end{equation}
The QCRB depends only on the reciprocal of the fluctuations of the generator $S_{GG}$.

This can be explained by the generalised uncertainty principle for waveform-estimation~\cite{miao2017general}
\begin{equation}\label{eq:cts GHUP}
    S_{x_\theta x_\theta}(\Om)S_{GG}(\Om) \geq \abs{\chi_{x_\theta G}(\Om)}^2 + \abs{S_{x_\theta G}(\Om)}^2 + c(\Om)
\end{equation}
where, in the quantum limit, this is an equality and $c(\Om)$ is a term that vanishes~\cite{Miao+2017}.

From Eq.~\ref{eq:cts GHUP}, Ref.~\cite{Miao+2017} showed that measuring the stationary complex quadrature $\h x_\theta(\Om)$ saturates the QCRB in Eq.~\ref{eq:waveform QCRB} if and only if it is uncorrelated with the generator, i.e.\ $S_{x_\theta G}=0$. Furthermore, Ref.~\cite{Miao+2017} showed that when $S_{x_\theta G}\neq0,\, \forall\theta$ the optimal error $S_{ss}$ is still within a factor of two of the QCRB.

It is not necessary a priori, however, that we measure a stationary complex quadrature $\h x_\theta(\Om)$. Eq.~\ref{eq:waveform QCRB} only applies to such stationary measurements. This leaves several important questions unanswered. What is the QCRB in general and when can it be saturated? If it cannot be saturated, then what is the optimal precision? And, what measurement attains this limit? We will answer these questions and demonstrate our results using gravitational-wave interferometry.

\textbf{\textit{Cosine and sine phases}}---At the positive frequency $\Om$, let $\tilde{s}(\Om)=\pi T(A + iB)$ where $A$ and $B$ are independent real degrees-of-freedom and $T$ is the finite integration time. In the time domain, this component of the signal is $A\cos(\Om t) + B\sin(\Om t)$ at a given time $t$ where $A$ and $B$ are the cosine and sine phases of the signal $s(t)$ at frequency $\Om$, respectively. 

Our goal of measuring the signal $s$, therefore, is equivalent to simultaneously estimating $A$ and $B$ at each $\Om$.

Our weighted figure-of-merit for the precision at $\Om$ is $\Si = 2w\text{Var}[\h A] + 2(1-w)\text{Var}[\h B]$ where $w\in(0,1)$ and $\h A$ and $\h B$ are unbiased estimates of $A$ and $B$, respectively. (Without loss of generality, we can assume that our weight matrix is diagonal.) The weights may be unequal $w\neq0.5$ for several reasons. For example, if we want to estimate the signal's power $\abs{\tilde{s}}^2\propto A^2+B^2$ more than its phase, then the weights would be unequal because the derivatives of $\abs{\tilde{s}}^2$ depend on $A$ and $B$.
\changed{We assume a uniform prior on $A$ and $B$ and distinguish that, while unequal weights indicate how much more is wanted to be known a posteriori about $A$ than $B$, a non-uniform prior would indicate how much more is known a priori about $A$ than $B$~\cite{van2004detection,gill1995applications,Genoni+2013}.}

Similarly to the signal, we can split the complex quadratures of the outgoing light $\h x_\theta(\Om)$ into their real and imaginary parts in the frequency domain. Or, equivalently, into their cosine and sine phases in the time domain, e.g.\ see Refs.~\cite{Branford+18, gefen2022quantum}. These parts are Hermitian but \textit{nonstationary}. In this manner, Eq.~\ref{eq:IOrelation-FD} becomes
\begin{equation}\label{eq:IOrelation-FD-parts}
    \vec{\h q} := \frac{1}{\sqrt{\pi T}}\bmatrixByJames{
        \re{\h x(\Omega)} \\ \re{\h p(\Omega)} \\ \im{\h x(\Omega)} \\ \im{\h p(\Omega)}
    }
    = \vec{\h q}^{\,(0)} + A \vec{d}_A + B \vec{d}_B.
\end{equation}
Let $\vec{\chi}=(\chi_{xG}(\Om), \chi_{pG}(\Om))^\T$, then the real signal displacements are
\begin{equation}\label{eq:signal displacements}
    \vec{d}_A := \sqrt{\pi T} \bmatrixByJames{
    \re{\vec{\chi}\!\;}
    \\ \im{\vec{\chi}\!\;}
        }, \quad
    \vec{d}_B := \sqrt{\pi T} \left[\!\begin{array}{r}
     -\im{\vec{\chi}\!\;}  
    \\ \re{\vec{\chi}\!\;} 
    \end{array}\right].
\end{equation}
These are orthogonal and have the same Euclidean norm

\begin{equation}\label{eq:l}
    l = \sqrt{\pi T\left(\abs{\chi_{x G}(\Om)}^2 + \abs{\chi_{p G}(\Om)}^2\right)}.
\end{equation}
Since $[\h x(\Om), \h p(\Om')] = i 2\pi\delta(\Om-\Om')$, by using $\re{z}=\frac{1}{2}(z + z^*)$ and $\im{z}=\frac{1}{2i}(z - z^*)$ we have that $[{\vec{\h q}}_1, {\vec{\h q}}_2]=[{\vec{\h q}}_3, {\vec{\h q}}_4]=i$ with all other commutators zero (${\vec{\h q}}_j$ is the $j$th element of $\vec{\h q}$). The system at each frequency, therefore, comprises two real displaced harmonic oscillators.

We assume that the noise is stationary such that the complex quadratures $\vec{\h x}=(\h x(\Om), \h p(\Om))^\T$ have the 2-by-2 covariance matrix $\frac{1}{2}\langle\{\vec{\h x}^{\,(0)}_j, \vec{\h x}^{\,(0)}_k\}\rangle=(V_2)_{jk}$ and the parts $\vec{\h q}$ have the 4-by-4 covariance matrix $\frac{1}{2}\langle\{\vec{\h q}^{\,(0)}_j, \vec{\h q}^{\,(0)}_k\}\rangle=(V_2\oplus V_2)_{jk}$. (Without loss of generality, we assume that $\langle{\vec{\h x}^{\,(0)}}\rangle=0$.) Since the device is linear, distinct frequencies are uncorrelated. For the moment, we assume that the pure state at each frequency is vacuum, i.e.\ $V_2=\text{diag}\left(\frac{1}{2},\frac{1}{2}\right)$, and will generalise later.

To measure $A$ and $B$ from the output light in Eq.~\ref{eq:IOrelation-FD-parts}, the na\"ive optimal unbiased estimates are $\h A_\text{na\"ive} = l^{-2}\vec{d}_A\cdot\vec{\h q}$ and $\h B_\text{na\"ive} = l^{-2}\vec{d}_B\cdot\vec{\h q}$. The variance of each estimate is $\frac{1}{2}l^{-2}$ such that the na\"ive figure-of-merit is $\Si_\text{na\"ive}=l^{-2}$. 
\changed{(Note that $\Si_\text{na\"ive}\propto T^{-1}$ by Eq.~\ref{eq:l} such that integrating for longer times reduces the error as expected.)}

These measurements, however, may not commute since $[\h A_\text{na\"ive}, \h B_\text{na\"ive}]= i\mu l^{-2}$ where
\begin{equation}\label{eq:mu}
    \mu = 2\pi Tl^{-2}\left(\re{\chi_{pG}}\im{\chi_{xG}}-\re{\chi_{xG}}\im{\chi_{pG}}\right)
\end{equation}
such that $0\leq\abs{\mu}\leq1$. (Without loss of generality, we assume that $\mu\geq0$.) 
This means that $A$ and $B$ cannot be simultaneously estimated to attain the na\"ive figure-of-merit if $\mu\neq0$. The displacements in Eq.~\ref{eq:signal displacements} are generated by their conjugate quadratures $\h G_A$ and $\h G_B$ which obey the same commutation relation such that $\mu=0$ is equivalent to the weak commutativity condition $\langle[\h G_A, \h G_B]\rangle=0$. The QCRB for simultaneous estimation of $A$ and $B$, therefore, can be saturated if and only if $\mu=0$ which is equivalent to $\exists\theta\in\R$ such that $\chi_{x_\theta G}=0$. 
\changed{In fact, the QCRB is precisely the na\"ive figure-of-merit above such that $\Si\geq\Si_Q=\Si_\text{na\"ive}$. This can be shown from the result that the QCRB with respect to $s_j$ given the unitary transformation $\exp(-i s_j \h G_j)$ is $(4 \var{G_j})^{-1}$~\cite{braunstein1994statistical}.}

\textbf{\textit{Fundamental precision limit}}---If $\mu\neq0$ such that the QCRB cannot be saturated, then the optimal attainable precision is instead the Holevo Cram\'er-Rao Bound (HCRB) $\Sigma_H$ which accounts for the commutator of the estimates such that $\Si\geq\Si_H>\Si_Q$~\cite{Holevo2011book,Genoni+2013,Bradshaw+2017,Bradshaw+2018}. Since the real parameters $A$ and $B$ appear as the shift in the mean of a pure Gaussian state, the HCRB is saturated by the optimal commuting linear combinations of $\vec{\h q}$~\cite{Holevo2011book}. (This is equivalent to finding the optimal Quantum Mechanics--free subspace~\cite{tsang2012evading}.) 
We calculate the HCRB using the method from Ref.~\cite{Bradshaw+2017} in the Supplemental Material~\cite{SupplementalMaterial}.

\begin{figure}[ht]
    \centering
    \includegraphics[width=\columnwidth]{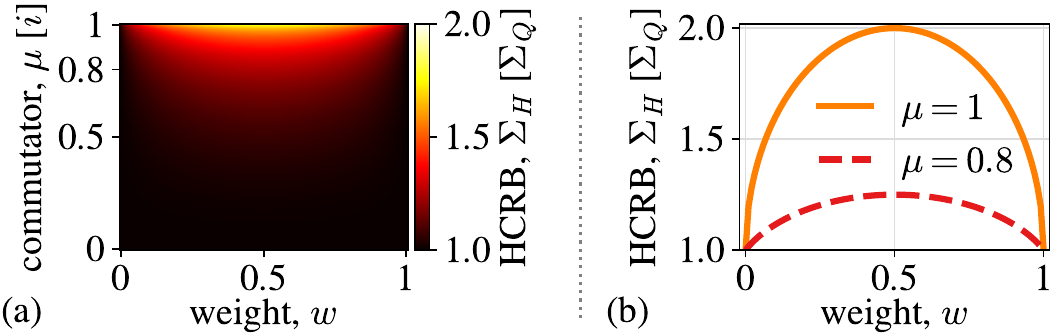}
    \caption{(a) HCRB versus the commutator and weight. (b) HCRB versus the weight for different commutators.} 
    \label{fig:2}
\end{figure}

We show that the ratio of the HCRB $\Si_H$ to the QCRB $\Si_Q=l^{-2}$ reduces to single-parameter optimisation
\begin{align}
    \label{eq:HCRB-semianalytic}
    \frac{\Si_H}{\Si_Q}
    &= \min_{\phi\in(0,\pi]} \left(
    \frac{w}{\cos(\phi)^2}
    + \frac{1-w}{\cos(\phi+\arcsin(\mu))^2}
    \right)
        \intertext{and we find analytic solutions in certain limits}
    \frac{\Si_H}{\Si_Q} \overset{\mu=1}&{\rightarrow} 1 + 2\sqrt{w(1 - w)}
    , \;
    \frac{\Si_H}{\Si_Q}
    \overset{w=\frac{1}{2}}{\rightarrow} \frac{2}{1+\sqrt{1-\mu^2}}
    .
\end{align}
Fig.~\hyperref[fig:2]{2a} shows that the ratio of the HCRB to the QCRB is at most two which agrees with Ref.~\cite{Miao+2017}. The HCRB increases monotonically with $\mu$ and decreases as the weights become less equal as shown in Fig.~\hyperref[fig:2]{2b}. The HCRB reduces to the QCRB for single-parameter estimation at $w=0,1$. 

These results generalise to squeezed states. Let $\h S$ be the conjugate squeezing operator such that $V_2\mapsto \text{diag}\left(\frac{1}{2},\frac{1}{2}\right)$. This unitary transformation does not affect $\mu$ or the bounds but does map the signal displacements as $\vec d\mapsto (S \oplus S)\vec d$ such that 
\begin{equation}\label{eq:lprime}
    l \mapsto l' := \sqrt{\pi T\left(\norm{S\re{\vec{\chi}\!\;}}^2 + \norm{S\im{\vec{\chi}\!\;}}^2\right)}.
\end{equation}

The general stationary pure Gaussian state case, therefore, is equivalent to the vacuum case with the same $\mu$ and $w$ but with $\Si_Q = (l')^{-2}$. We emphasise that we only apply $\h S$ mathematically to derive the bounds; it is not required experimentally.

\textbf{\textit{Optimal measurement scheme}}---There exists a unique symplectic transformation of the two harmonic oscillators $\vec{\h q}\mapsto\vec{\h X}=(\h X_1, \h P_1, \h X_2, \h P_2)^\text{T}$ that maps the normalised displacements as $l^{-1}\h d_A\mapsto\h X_1$ and $l^{-1}\h d_B\mapsto\mu\h P_1 + \sqrt{1-\mu^2}\h X_2$ such that their commutator remains $i\mu$. In this basis, the optimal commuting unbiased estimates are~\cite{SupplementalMaterial}
\begin{align}\label{eq:optimal estimates}
    \h A &= [l\cos(\barBetter{\phi})]^{-1}(\cos(\barBetter{\phi}) \h X_1 - \sin(\barBetter{\phi}) \h P_2) \\
    \nonumber
    \h B &= [l\cos(\barBetter{\phi}+\arcsin(\mu))]^{-1}(\cos(\barBetter{\phi}) \h X_2 - \sin(\barBetter{\phi}) \h P_1) 
\end{align} 
where $\barBetter{\phi}$ is the optimal angle in Eq.~\ref{eq:HCRB-semianalytic}. 

These estimates are two nonstationary quadratures: arbitrary real linear combinations of $\vec{\h q}$. Compare this to the stationary complex quadrature $\h x_\theta(\Om)$ with the real part $\cos(\theta){\vec{\h q}}_1 + \sin(\theta){\vec{\h q}}_2$ and imaginary part $\cos(\theta){\vec{\h q}}_3 + \sin(\theta){\vec{\h q}}_4$.
For squeezed states, similarly, the optimal measurement consists of two nonstationary quadratures (mathematically, first apply $\h S$ and then the symplectic transformation).

\begin{figure}[t]
    \includegraphics[width=\columnwidth]{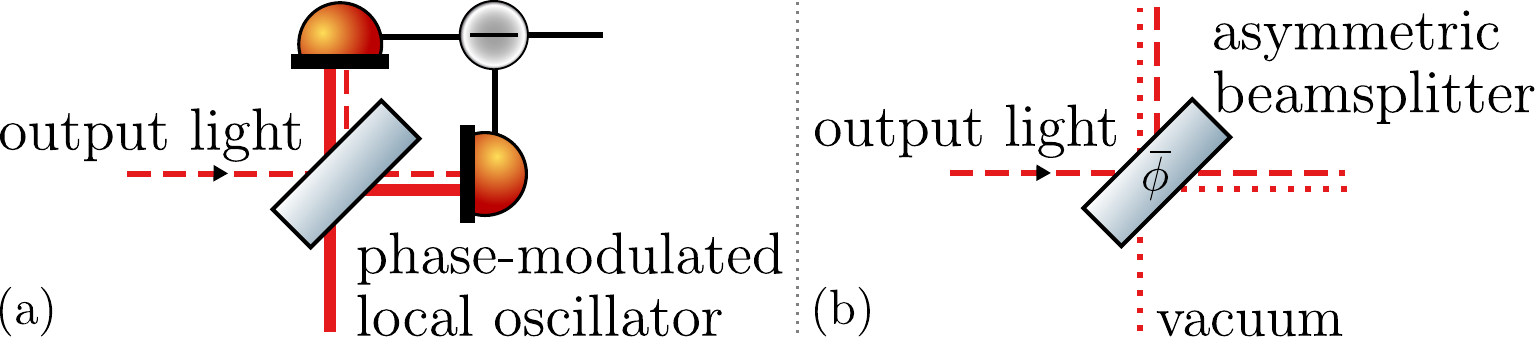}
    \caption{(a) Phase-modulated balanced homodyne readout. (b) Asymmetric beamsplitter with power reflectivity $\cos(\bar\phi)^2$. 
    }
    \label{fig:3}
\end{figure}

We propose how to experimentally realise these nonstationary measurements at a given $\Omega$. We expand $\vec{\h q}$ into the time domain using $\re{\h x_\theta(\Om)} = \intginf{t} \cos(\Om t) \h x_\theta(t)$ and $\im{\h x_\theta(\Om)} = \intginf{t} \sin(\Om t) \h x_\theta(t)$. Since the measurements are linear combinations of $\vec{\h q}$, therefore, they are $\h A = \intginf{t} c_A(t) \h x_{\theta_A(t)}(t)$ and $\h B = \intginf{t} c_B(t) \h x_{\theta_B(t)}(t)$ for some real amplitudes $c_A(t)$ and $c_B(t)$ and phases $\theta_A(t)$ and $\theta_B(t)$~\cite{SupplementalMaterial}. For example, if $w=1$ such that we only want to estimate $A$, then, as shown in Fig.~\hyperref[fig:3]{3a}, we use homodyne readout with a phase-modulated local oscillator with phase $\theta_A(t)$ to obtain the timeseries $\h x_{\theta_A(t)}(t)$. By integrating this timeseries multiplied by $c_A(t)$ in post processing, we can achieve $\h A$.

Suppose that instead we want to measure both $\h A$ and $\h B$. Although $\h A$ and $\h B$ commute, their integrands $c_A(t) \h x_{\theta_A(t)}(t)$ and $c_B(t) \h x_{\theta_B(t)}(t)$ above may not commute at a given time. This prevents directly performing simultaneous modulated homodyne measurements. If $\mu=1$ such that the normalised displacements are $\h X_1$ and $\h P_1$, however, then this can be overcome by using an asymmetric beamsplitter with reflectivity $\cos(\bar\phi)^2$ to mix in an ancillary mode (i.e.\ uncorrelated vacuum) as shown in Fig.~\hyperref[fig:3]{3b}. Then, measuring $\vec{d}_A$ and $\vec{d}_B$ on the two output modes from the beamsplitter using modulated homodyne readouts commutes at each time and saturates the HCRB for any $w$~\cite{SupplementalMaterial}. The added noise from the ancilla is responsible for the gap from the QCRB.

For the general case of any $\mu$ and $w$, we propose a joint homodyne-heterodyne readout scheme. To obtain the individual estimate of $\h A$ above, we integrated the timeseries $c_A(t)\h x_{\theta_A(t)}(t)$, but information is available at other frequencies too. In particular, the $2\Om$ Fourier component beats with the timeseries which oscillates at $\Om$ to produce a linear combination of the quadratures at $\Om$ and $3\Om$. This can realise a heterodyne measurement of $\h B$ at $\Om$~\cite{buonanno2003quantum,SupplementalMaterial}. The added heterodyne noise at $3\Om$ can be suppressed by squeezing the output mode using two cascaded, detuned, and narrowband filter cavities---one for each of the upper and lower sidebands at $3\Om$---without affecting the estimates or the fundamental limits at $\Om$. The HCRB, therefore, can be saturated in the narrowband around $\Om$ using a homodyne measurement of $\h A$ and a simultaneous heterodyne measurement of $\h B$.

\begin{figure}
    \centering
    \includegraphics[width=\columnwidth]{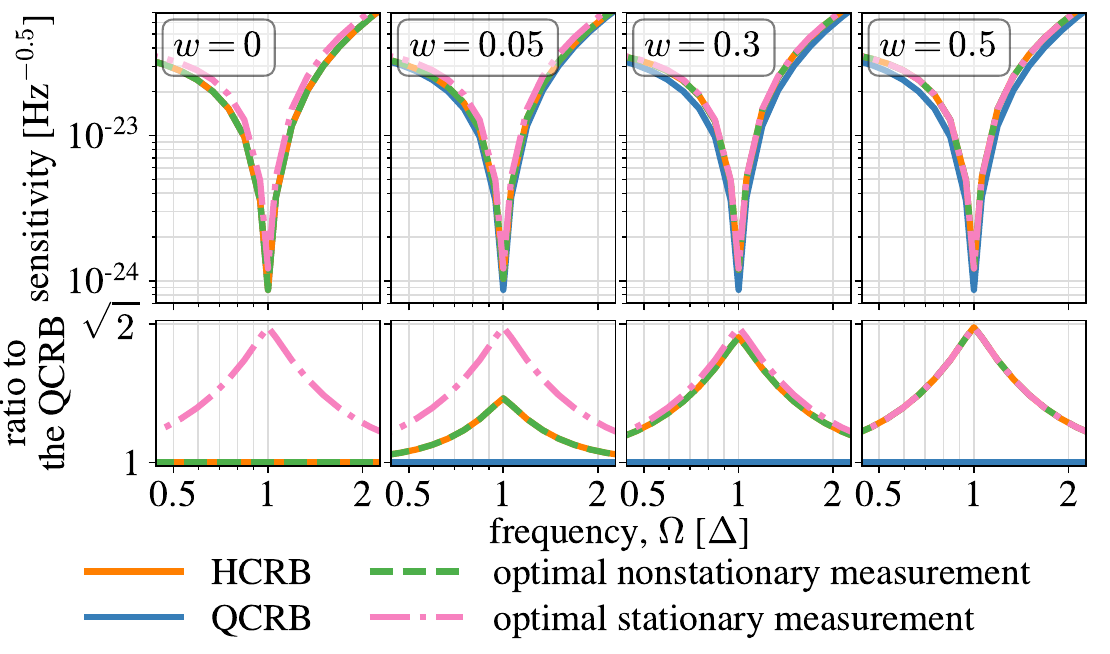}
    \caption{Strain sensitivity for the detuned LIGO-like interferometer versus frequency for different weights in (top row) effective amplitude spectral density units and (bottom row) ratio to the QCRB.}
    \label{fig:4}
\end{figure}
\textbf{\textit{Gravitational-wave interferometry.}}---We demonstrate our results for a gravitational-wave interferometer like the Laser Interferometric Gravitational-wave Observatory (LIGO)~\cite{AdvancedLIGO:2015,buikemaSensitivityPerformanceAdvanced2020} operated in a hypothetical detuned configuration. For simplicity, we model LIGO as a power-recycled Fabry-P\'erot Michelson interferometer as shown in Fig.~\hyperref[fig:1]{1b} with vacuum input into the ``dark port'' of the beamsplitter~\cite{meersRecyclingLaserinterferometricGravitationalwave1988,bond_2010}. In our detuned configuration, the 4-km arm cavities with 750~kW of circulating power are detuned away from the input carrier laser frequency of 282~THz by $\Delta=2\pi\times3$~kHz~\cite{PhysRevD.103.022002,ward2010length,somiya2012detector,miyakawa2006measurement,buonanno2003scaling,corbitt2004quantum,Ganapathy_2021}. We are interested in detecting 1--4~kHz gravitational-wave signals, e.g.\ from the post-merger remnant of binary neutron-star mergers to test our theories of extreme matter~\cite{lasky_2015,BAIOTTI2019103714,bauswein2019identifying,Universe7040097,PhysRevD.79.044030,Ott_2009,PhysRevX.4.041004,postmerger_long,postmerger_short}. (Since we focus on the kilohertz response, we ignore quantum radiation pressure noise~\cite{danilishinQuantumMeasurementTheory2012}.) Detuning the interferometer makes it resonant at $\Delta$ which improves the peak sensitivity without increasing the circulating power~\cite{Brooks_2021,PhysRevLett.114.161102,Barsotti_2018}.
We emphasise that operating LIGO in a detuned configuration presents many technical challenges~\cite{Ganapathy_2021} and here we only want to establish the fundamental limit of achievable sensitivity at a given frequency to better evaluate this configuration.

The differential optical mode of the interferometer can be approximated as a single mode in a detuned cavity linearly coupled to the gravitational-wave strain $s(t)$ by $\h H_\text{int} = g s(t) \h x_\text{cav}$. Here, $g$ is the effective coupling rate (mediated by free masses in the transverse-traceless gauge) and $\h x_\text{cav}$ is the amplitude quadrature of the intracavity mode such that $\h G = -g \h x_\text{cav}$~\cite{kimble_2001}. The resulting susceptibility is~\cite{SupplementalMaterial}
\begin{equation}
    \vec{\chi} = \frac{\sqrt{2\gamma} g}{\Delta ^2+(\gamma -i \Omega )^2} 
    \bmatrixByJames{\Delta \\ -\gamma +i \Omega} 
\end{equation}
where $\gamma=2\pi\times42$~Hz is the half-width at half-maximum readout rate of the arm cavities. By Eq.~\ref{eq:mu}, $\mu=\frac{2\Delta\Omega }{\gamma ^2+\Delta ^2+\Omega ^2}$ such that the QCRB cannot be saturated for $\Delta\neq0$ which agrees with Ref.~\cite{Miao+2017}. 

In Fig.~\ref{fig:4}, we compare the HCRB versus frequency to the sensitivity using the optimal stationary quadrature (also known as ``variational readout'') and nonstationary quadrature measurements. For equal weights, the stationary measurement saturates the HCRB such that the gap to the QCRB is insurmountable. For unequal weights, however, our nonstationary measurement is required to saturate the HCRB.

This unequal weight regime is relevant because, e.g., astrophysically we care more about knowing some parameters of the neutron-star equation-of-state than others. This can be reduced to having an unequal weighting between the signal's power and phase and, therefore, between $A$ and $B$.
\changed{For example, we particularly want to estimate the primary peak of the kilohertz power spectrum to inform our understanding of the equation-of-state~\cite{topolski2023post}. In the future, with more precise numerical models of the post-merger signal, we may be able to confidently determine the phase of the post-merger signal from a strong enough detection of the inspiral phase. We then only need to estimate the post-merger signal's power which is equivalent to having weights $w=0$ or $1$. In this limiting single-parameter case, we have shown that} our nonstationary measurement scheme improves the signal-to-noise ratio by up to a factor of $\sqrt2$ at the detuning frequency, an improvement which cannot be surpassed using a different measurement scheme. This corresponds to up to a factor of $2.83$ improvement in the volume of the Universe searched for kilohertz signals at the peak frequency, in addition to the gain provided from detuning the interferometer. This could be a significant boost to LIGO's search for kilohertz gravitational-waves should the challenges with detuned interferometry be overcome.
\changed{More realistically, we may instead have partial prior knowledge of the post-merger signal's phase and perform weighted simultaneous estimation of the signal's power and phase. We hypothesise that the sensitivity can still be similarly improved in this regime and defer a detailed study of this application to future work.}

Losses limit the possible quantum enhancement of LIGO (where Fig.~\ref{fig:4} shows the lossless sensitivity). If we assume optical losses of 100~ppm ($\gamma_l=2\pi\times0.3$~Hz) in the arm cavities and $\eta=0.1$ in the output, then $l\mapsto\sqrt{1-\eta}l$ and $\mu\mapsto\frac{2\Delta\Omega }{(\gamma+\gamma_l) ^2+\Delta ^2+\Omega ^2}$~\cite{SupplementalMaterial}. Since $\gamma_l\ll\gamma\ll\Delta$, $\mu$ is unchanged. This implies that the gap from the HCRB to the relevant QCRB, $\Si_Q\approx(1-\eta)^{-1}l^{-2}$, is also unchanged and our nonstationary measurement can still achieve up to a factor of $\sqrt2$ improvement with losses.

\textbf{\textit{Conclusions}}---We have shown how to achieve the fundamental precision limit for the estimation of a classical signal using a linear quantum device. Previous work on linear waveform estimation found an unexplained gap of up to a factor of $\sqrt2$ in the signal-to-noise ratio between the optimal stationary quadrature measurement and the QCRB. We showed that this gap stems from the non-commutativity of the na\"ive estimates of the cosine and sine phases of the signal at each frequency. This allowed us to establish the fundamental limit of attainable precision and propose how to experimentally realise the optimal nonstationary measurement scheme.
We applied these results to the search for post-merger gravitational-wave signals from binary neutron-star mergers using a detuned LIGO-like interferometer. We showed that this nonstationary measurement scheme could significantly increase the volume of the Universe probed for such signals at a given frequency in the unequal weight regime.

Future work could determine the broadband optimal measurement scheme and apply our results to a dual-recycled LIGO-like interferometer with injected squeezed states~\cite{aasietal2013,tseQuantumEnhancedAdvancedLIGO2019,mcculler2020frequency} and extend them to other systems, e.g.\ $\mathcal{PT}$-symmetric interferometers~\cite{Wang+2022,Metelmann+2014,li2020broadband,liEnhancingInterferometerSensitivity2021,Bentley+2021arXiv,PhysRevD.106.L041101}, axion detectors~\cite{PhysRevLett.51.1415,PhysRevLett.124.101303,PhysRevX.9.021023,MARSH20161}, and displacement noise--free interferometers~\cite{Chen06a,Chen06b,gefen2022quantum}.

Our code is available online~\cite{FIrepo} and was written using \textsc{Mathematica}~\cite{mathematica} and \textsc{Python}~\cite{python,ipython,jupyter,numpy,matplotlib}.

\begin{acknowledgements}
We thank the following individuals and groups for their advice provided during this research: L.~McCuller, S.~Kotler, J.~Preskill, Z.~Mehdi, A.~Markowitz, D.~Ganapathy, S.~Zhou, L.~Sun, R.~X.~Adhikari, the Caltech Y.~Chen Quantum Group, and the ANU CGA Squeezer Group. We use component graphics from Ref.~\cite{ComponentLibrary} with permission. This research is supported by the Australian Research Council Centre of Excellence for Gravitational Wave Discovery (Project No. CE170100004). J.W.G. and this research are supported by an Australian Government Research Training Program (RTP) Scholarship and also partially supported by the US NSF grant PHY-2011968. In addition, Y.C. acknowledges the support by the Simons Foundation (Award Number 568762). T.G. acknowledges funding provided by the Institute for Quantum Information and Matter and the Quantum Science and Technology Scholarship of the Israel Council for Higher Education. S.A.H. acknowledges support through an Australian Research Council Future Fellowship grant FT210100809.
This Letter has been assigned LIGO Document No.\ P2300096.
\end{acknowledgements}

\end{document}

\title{Supplemental Material for \\\emph{Holevo Cram\'er-Rao Bound for waveform estimation of gravitational waves}}

\author{James~W.~Gardner}
\email{james.gardner@anu.edu.au}

\author{Tuvia Gefen}
\email{tgefen@caltech.edu}

\author{Simon A. Haine}
\author{Joseph J. Hope}

\author{Yanbei Chen}

\date{\today}
\maketitle
\allowdisplaybreaks

In this Supplemental Material, we present the following details for completeness and further clarity.

\tableofcontents

\section{The detuned interferometer}
This section accompanies the introductory section in the Letter.

\subsection{Review of the two-photon formalism}
We work within the framework of the two-photon formalism~\cite{caves1985new}. Let us therefore review this formalism and formally define what we mean in the Letter by a quadrature at a given frequency.\\

We start by quantising the electric field. Let $\h a'(t)$ be the ``one-photon'' (also called the ``single-photon'') annihilation operator for the electric field at time $t$~\cite{caves1985new,danilishinQuantumMeasurementTheory2012}. Let the Fourier transform of some quantity $Q$ at frequency $\Omega$ be
\begin{align}\label{eq:FourierTransformDef}
    Q(\Omega)&= \intginf{t} e^{i\Omega t} Q(t)
    \suchthat
    Q(t)&= \frac{1}{2\pi}\intginf{\Om} e^{-i\Omega t} Q(\Om)
    .
\end{align}
Then, we can define the frequency-domain annihilation operator for $\Omega \geq 0$ as
\begin{align}
    \h a'(\Om) = \intginf{t} e^{i\Omega t} \h a'(t)
\end{align}
and $\h a'^{\dagger}\left(\Omega\right)$ as its Hermitian conjugate. We define the one-photon quadrature at quadrature angle $\theta$ as
\begin{align}
    \h x'_\theta(t) &= \frac{1}{\sqrt2}\left(e^{-i\theta}\hat a'(t) + e^{i\theta}\hat a'^\dag(t)\right). \\
    \intertext{The frequency-domain one-photon quadratures are defined analogously as}
    \h x'_{\theta,\Om} &= \frac{1}{\sqrt2}\left(e^{-i\theta}\hat a'(\Om) + e^{i\theta}\left[\hat a'(\Om)\right]^\dag\right)
    .
\end{align}
These quadratures are the standard Hermitian quadratures of the harmonic oscillator that corresponds to the frequency $\Omega$ mode of the electromagnetic field.\\

In the ``two-photon'' formalism, we factor out oscillations at a given frequency $\om$ (e.g.\ at the carrier frequency $\om_0$ of the laser of the interferometer) such that the two-photon annihilation operator $\h a(t)$ in terms of the one-photon annihilation operator $\h a'(t)$ is
\begin{align}
    \h a(t)&=e^{i\om t}\h a'(t)
    \intertext{and similarly}
    \h a^\dag(t)&=e^{-i\om t}\h a'^\dag(t)
\end{align}    
such that the Fourier transforms of the annihilation operators are related by
\begin{align}
\label{eq:two_photons_1}
    \h a(\Om) &=\intginf{t} e^{i\Omega t} \h a(t) \\
    &=\intginf{t} e^{i(\om+\Omega) t} \h a'(t)\\
    &=\h a'(\omega+\Om).
\end{align}
Similarly, their frequency-domain Hermitian conjugates are related by

\begin{align}
    \hat{a}^{\dagger}\left(\Omega\right)
    &=\intginf{t}e^{i\Omega t}\hat{a}^{\dagger}\left(t\right)\\
    &=\intginf{t}e^{i\left(\Omega-\omega\right)t}\hat{a}'^{\dagger}\left(t\right)
    \\&=\left[\h a'\left(\omega-\Omega\right)\right]^\dag.
\label{eq:two_photons_2}
\end{align}
Here, we now have that $-\omega\leq \Omega \leq \omega,$ and by taking the approximation of large $\omega$ we can take the integration range to be from $-\infty$ to $\infty$.
The ``two-photon'' quadrature at angle $\theta$ in the frequency-domain, therefore, is
\begin{align}
\h x_\theta(\Om) &= \frac{1}{\sqrt{2}}\left(e^{-i \theta}\hat{a}\left(\Omega\right)+e^{i \theta}\hat{a}^{\dagger}\left(\Omega\right)\right) \\
\intertext{which, using Eq.~\ref{eq:two_photons_1} and Eq.~\ref{eq:two_photons_2}, becomes}
\h x_\theta(\Om)&=\frac{1}{\sqrt2}\left(e^{-i\theta}\hat a'(\om+\Om) + e^{i\theta}\left[\h a'\left(\omega-\Omega\right)\right]^\dag\right).
\end{align}
The two-photon quadratures, therefore, combine the one-photon quadratures at two frequencies $\om\pm\Om$. For example,
\begin{align}
    \h x(\Om) &= \frac{1}{\sqrt2}\left(\hat a'(\om+\Om) + \left[\h a'\left(\omega-\Omega\right)\right]^\dag\right) \\
    &= \frac{1}{2}\left(\hat x'_{\om+\Om} + \hat x'_{\om-\Om}\right) + \frac{i}{2}\left(\hat p'_{\om+\Om}- \hat p'_{\om-\Om}\right)
    .
\end{align}
With $\omega=\omega_0$, these two-photon quadratures are what we implicitly use in the Letter.\\

The real and imaginary parts of the two-photon quadratures are therefore
\begin{align}\label{eq:OnePhotonFormalism}
    \vec{\h q}^{(\omega)}(\Om) &= \frac{1}{2\sqrt{\pi T}}
    \bmatrixByJames{
    1 & 0 & 1 & 0 \\
    0 & 1 & 0 & 1 \\
    0 & 1 & 0 & -1 \\
    -1 & 0 & 1 & 0 \\
    }\bmatrixByJames{
        \h x'_{\om+\Om} \\
        \h p'_{\om+\Om} \\
        \h x'_{\om-\Om} \\
        \h p'_{\om-\Om} \\
    }
\end{align}
where we indicate which ``centre frequency'' $\omega$ the two-photon parts $\vec{\h q}^{(\omega)}$ are defined with respect to. (We implicitly work with $\vec{\h q}^{(\omega_0)}$ in the Letter, e.g.\ in Eq.~4.) This transformation is the same as 
\begin{align}
    \bmatrixByJames{
    \cos(\om t)\cos(\Om t) \\
    \sin(\om t)\cos(\Om t) \\
    \cos(\om t)\sin(\Om t) \\
    \sin(\om t)\sin(\Om t) \\
    }
    =
    \frac{1}{2}
    \bmatrixByJames{
    1 & 0 & 1 & 0 \\
    0 & 1 & 0 & 1 \\
    0 & 1 & 0 & -1 \\
    -1 & 0 & 1 & 0 \\
    }
    \bmatrixByJames{
    \cos((\om+\Om) t) \\
    \sin((\om+\Om) t) \\
    \cos((\om-\Om) t) \\
    \sin((\om-\Om) t) \\
    }.
\end{align}

If we shift by $\delta\om=\om'-\om$ between describing the light using the two-photon formalism centred at $\omega$ and using the two-photon formalism centred at $\omega'$, then by Eq.~\ref{eq:OnePhotonFormalism} the parts transform as
\begin{align}
    \label{eq:Shift2PhotonParts}
    &
    \vec{\h q}^{(\om')}(\Om)
    = \text{M}_\text{shift}
    \bmatrixByJames{
        \vec{\h q}^{(\om)}(\delta\om+\Om) \\
        \vec{\h q}^{(\om)}(\delta\om-\Om) \\
    }
    \where
        &\text{M}_\text{shift}
    =
    \frac{1}{2}
    \bmatrixByJames{
        1 & 0 & 0 & -1 & 1 & 0 & 0 & -1 \\
         0 & 1 & 1 & 0 & 0 & 1 & 1 & 0 \\
         0 & 1 & 1 & 0 & 0 & -1 & -1 & 0 \\
         -1 & 0 & 0 & 1 & 1 & 0 & 0 & -1 \\
    }
    .
\end{align}
This transformation will be useful later.

\subsection{Hamiltonian model of the detuned interferometer}

We model the detuned FPMI's differential mode as a single mode at the carrier frequency $\omega_0$ inside of an optical cavity detuned at frequency $\Delta$. As described in the Letter, we use the single-mode and free-mass approximations which are valid when studying the 1--4~kHz response of the interferometer. Let the annihilation (creation) operator of the cavity mode be $\hat a^\text{(cav)}$ ($\hat a^{\text{(cav)}^\dag}$). The intracavity optical quadratures are parameterised by a quadrature angle $\theta$ as
\begin{align}\label{eq:quadratureDef}
    \hat x_\theta^\text{(cav)}
    & = \frac{1}{\sqrt2}\left(e^{-i\theta}\hat a^\text{(cav)} + e^{i\theta}\hat a^{\text{(cav)}^\dag}\right)\\
    &= \cos(\theta)\hat x^\text{(cav)} + \sin(\theta)\hat p^\text{(cav)} \\
    \suchthat
    \hat x^\text{(cav)} &= \hat x_0^\text{(cav)},\quad \hat p^\text{(cav)} = \hat x_{\frac{\pi}{2}}^\text{(cav)}
    .
\end{align}
The nonzero commutators of these operators are
\begin{align}
    [\hat a^\text{(cav)}, \hat a^{\text{(cav)}^\dag}] &= 1, \quad
    [\hat x^\text{(cav)}, \hat p^\text{(cav)}] = i
    .
\end{align}

Let the continuum of modes $\hat a^\text{(bath)}(\omega)$ outside the cavity be indexed by their frequency $\omega$. These bath modes define the vacuum mode $\hat a^\text{(in)}$ incident on the dark port of the interferometer and the reflected mode $\hat a$ containing the gravitational-wave signal and measured at the photodetector~\cite{gardiner1985input}. ($\hat a$ is commonly also notated as $\hat a^\text{(out)}$.) Similarly to Eq.~\ref{eq:quadratureDef}, these annihilation operators define quadratures $\hat x_\theta^\text{(bath)}$, $\hat x_\theta^\text{(in)}$, and $\hat x_\theta$.

These different quadratures are connected via the input/output relation
\begin{align}\label{eq:IOrelation}
    \hat x_\theta = - \hat x_\theta^\text{(in)} + \sqrt{2 \gamma} \hat x^\text{(cav)}_\theta.
\end{align}
Here, $\gamma = -c/(4L)\log(1-T)$ is the cavity HWHM bandwidth where $L$ is the arm cavities' length and $T$ is the input test masses' power transmissivity. \\

The Hamiltonian $\hat H$ describing the detuned interferometer is

\begin{align}\label{eq:Ham}
    \hat H &= \hat H_0 + \hat H_\text{det} + \hat H_\text{int} + \hat H_\text{I/O}\\
    \intertext{where}
    \hat H_0 &= \hbar\omega_0 \hat a^{\text{(cav)}^\dag} \hat a^\text{(cav)}\\ 
    \hat H_\text{det} &= -\hbar \Delta \hat a^{\text{(cav)}^\dag} \hat a^\text{(cav)}\\ 
    \hat H_\text{int} &= g h(t) \hat x^\text{(cav)}\\
    \frac{\hat H_\text{I/O}}{\hbar \sqrt{2\gamma}} &= \int_{-\infty}^\infty \text{d}\omega\, \biggl( i \hat a^{\text{(bath)}^\dag}(\omega)\hat a^\text{(cav)} + \text{H.c.} \biggr).
\end{align}
Here, $\hat H_0$ gives the free evolution, $\hat H_\text{det}$ gives the detuning, $\hat H_\text{int}$ gives the effective interaction between the gravitational-wave signal and the optical field mediated by free masses in the transverse-traceless gauge, and $\hat H_\text{I/O}$ gives the input/output relation in Eq.~\ref{eq:IOrelation}. We omit the free evolution of the bath modes for brevity. $\text{H.c.}$ is the Hermitian conjugate. The effective coupling rate $g$ in $\hat H_\text{int}$ has units of energy and is given by $c g^2 = 2 \hbar \omega_0 L P$ where $c$ is the speed of light and $P$ is the circulating power~\cite{li2020broadband}. 
We emphasise that, in this effective model, the mechanical evolution of the test masses is ignored and they are treated as having infinite mass. \\

In the Interaction Frame with respect to $\hat H_0$ and the free evolution of the bath modes, the Heisenberg-Langevin equations-of-motion derived from Eq.~\ref{eq:Ham} are
\begin{align}\label{eq:TimeDomainIO}
    \partial_t \hat x^\text{(cav)} &= - \gamma \hat x^\text{(cav)} + \sqrt{2\gamma} \hat x^\text{(in)} - \Delta \hat p^\text{(cav)} \\
    \partial_t \hat p^\text{(cav)} &= - \gamma \hat p^\text{(cav)} + \sqrt{2\gamma} \hat p^\text{(in)} + \Delta \hat x^\text{(cav)} - \frac{g}{\hbar} h(t).\nonumber
\end{align}
Here, the detuning $\Delta$ mixes the quadratures such that some of the signal $h(t)$ ends up in $\hat x^\text{(cav)}$. 
Then, Eq.~\ref{eq:TimeDomainIO} for the two-photon quadratures in the Fourier domain defined by Eq.~\ref{eq:FourierTransformDef} is
\begin{align}\label{eq:FourierDomainIO}
    (\gamma-i\Omega) \hat x^\text{(cav)}(\Omega) &= \sqrt{2\gamma} \hat x^\text{(in)}(\Omega) - \Delta \hat p^\text{(cav)}(\Omega) \\
    (\gamma-i\Omega) \hat p^\text{(cav)}(\Omega) &= \sqrt{2\gamma} \hat p^\text{(in)}(\Omega) + \Delta \hat x^\text{(cav)}(\Omega) - \frac{g}{\hbar} \tilde{h}.\nonumber
\end{align}
Here, we approximate integrals of the following kind since the integration time $T$ is large
\begin{align}
    \intg{-\frac{T}{2}}{\frac{T}{2}}{t} e^{i\Om t}\hat x^\text{(cav)}(t) \approx \hat x^\text{(cav)}(\Omega).
\end{align}
These Fourier-domain quantities $Q$ are complex but obey $Q^\dag(\Omega)=Q(-\Omega)$ because their time-domain counterparts are real. We only consider positive frequencies such that there are two independent degrees-of-freedom at each frequency. As discussed in the Letter, these are the real and imaginary parts in the frequency domain or the cosine and sine phases in the time domain. Eq.~\ref{eq:FourierDomainIO} can be linearly solved and then Eq.~\ref{eq:IOrelation} used to find that
\begin{align}
\label{eq:ComplexIOVectors}
    \bmatrixByJames{\hat x(\Omega) \\ \hat p(\Omega)} &= \vec{d_h} \tilde h \hat 1 + \bmatrixByJames{\hat x^\text{(0)}(\Omega) \\ \hat p^\text{(0)}(\Omega)} \\
    \intertext{where the signal response is}
    \vec{d_h} &= \bmatrixByJames{d_0 \\ d_{\frac{\pi}{2}}} \\
    &= \frac{\sqrt{2\gamma} g}{\hbar  \left(\Delta ^2+(\gamma -i \Omega )^2\right)} 
    \bmatrixByJames{\Delta \\ -\gamma +i \Omega} \\
    \intertext{and the noise is}
    \bmatrixByJames{\hat x^\text{(0)}(\Omega) \\ \hat p^\text{(0)}(\Omega)} &= \text{M}_\text{(in)}^\text{(0)} \bmatrixByJames{\hat x^\text{(in)}(\Omega) \\ \hat p^\text{(in)}(\Omega)} \\
    \intertext{where}
    \text{M}_\text{(in)}^\text{(0)} &= \frac{1}{\Delta ^2+(\gamma -i \Omega )^2} \label{eq:Min0}
    \\&\cdot \bmatrixByJames{\gamma ^2-\Delta ^2+\Omega ^2 & -2 \gamma  \Delta  \\
    2 \gamma  \Delta  & \gamma ^2-\Delta ^2+\Omega ^2 \\}\nonumber
    .
\end{align}
This is written compactly in Eq.~2 where
\begin{align}
    d_\theta &= \cos(\theta)d_0 + \sin(\theta)d_{\frac{\pi}{2}}
    .
\end{align}
Here, the signal response $\abs{d_\theta}$ is resonantly improved around $\Delta$ where we assume that $\Delta\gg\gamma$. $\text{M}_\text{(in)}^\text{(0)}$ is such that the measurement's noise $\hat x_\theta^{(0)}$ is still at vacuum.

\subsection{The tuned interferometer}

For illustrative purposes, we consider the tuned case with $\Delta=0$ where $d_0 = 0$ and
\begin{align}
    d_{\frac{\pi}{2}} = \frac{-\sqrt{2\gamma} g}{\hbar  \left(\gamma -i \Omega \right)} 
    .
\end{align}
And, where the cavity simply gives a phase delay
\begin{align}
    \hat p^\text{(0)}(\Omega) = \frac{\gamma + i \Omega}{\gamma - i \Omega} \hat p^\text{(in)}(\Omega).
\end{align}
Then, the optimal protocol is to measure $\hat p(\Omega)$ from Fourier transforming the timeseries results $\hat p$. The optimal unbiased estimate of $\tilde h$ is
\begin{align}\label{eq:TunedEst}
    \hat h = \frac{\hat p(\Omega)}{d_{\frac{\pi}{2}}}, \quad \ev{\hat h} = \tilde h.
\end{align}
The real and imaginary parts of the signal corresponding to $A$ and $B$ can be estimated from the real and imaginary parts of $\hat p(\Omega)$, respectively. \\

The single-sided, symmetrised power spectral density $S_{Z_1,Z_2}$ for stationary random processes $\hat Z_1(\Omega)$ and $\hat Z_2(\Omega)$ is
\begin{align}\label{eq:PSDdef}
    \frac{1}{2}2\pi\delta(\Omega-\Omega')S_{Z_1,Z_2}(\Omega)&=\ev{\frac{1}{2}\{\hat Z_1(\Omega), \hat Z_2^\dag(\Omega')\}_+}\\
    \intertext{where the anti-commutator is}
    \{\hat Z_1(\Omega), \hat Z_2^\dag(\Omega')\}_+&=\hat Z_1(\Omega)\hat Z_2^\dag(\Omega')+\hat Z_2^\dag(\Omega')\hat Z_1(\Omega).
\end{align}
Here, $\delta$ is the Dirac delta distribution in the frequency domain with units of time such that $\delta(0)\sim T$ where $T$ is the integration time. Its counterpart is $\delta_t$ in the time domain.
If the random processes each have zero mean, then with $\Omega=\Omega'$
\begin{align}
    S_{Z_1,Z_2}(\Omega) &= \frac{1}{\pi T} \cov{\hat Z_1(\Omega)}{\hat Z_2^\dag(\Omega)}\\
    \intertext{and}
    S_{Z_1,Z_1}(\Omega) &= \frac{1}{\pi T} \var{\hat Z_1(\Omega)}
    .
\end{align}
This is the origin of the reoccurring factor of $\pi T$ in our results. The quantum shot noise measured in $\hat h$ from Eq.~\ref{eq:TunedEst} is described by the power spectral density
\begin{align}\label{eq:TunedPSD}
    S_{h,h} &= \frac{1}{\pi T} \var{\frac{\hat p(\Omega)}{d_{\frac{\pi}{2}}}} \\
    &= \frac{1}{\abs{d_{\frac{\pi}{2}}}^2} \\
    &= \frac{\hbar^2\left(\gamma^2 + \Omega^2\right)}{2\gamma g^2}
    .
\end{align}

\section{The gap from the sensitivity to the QCRB} 

This section accompanies the section titled \emph{The gap} in the Letter.\\

For the detuned interferometer, Ref.~\cite{Miao+2017} found a gap between the sensitivity $S_{h,h}$ of estimating $\tilde{h}$ and the waveform-estimation QCRB~\cite{Tsang+2011}
\begin{align}\label{eq:wfQCRB}
    S_Q^\text{(wave)} = \frac{\hbar^2}{S_{F,F}}.
\end{align}
Here, $S_{F,F}$ is the power spectral density from Eq.~\ref{eq:PSDdef} of the probe observable $\hat F=g \hat x^\text{(cav)}$ in $\h H_\text{int}$ of Eq.~\ref{eq:Ham} such that
\begin{align}\label{eq:wfQCRBwithLittleG}
    S_Q^\text{(wave)} = \frac{\hbar ^2 }{2 g^2}
    \frac{\left(\gamma ^2+(\Delta -\Omega )^2\right) \left(\gamma ^2+(\Delta +\Omega )^2\right)}{ \gamma\left(\gamma ^2+\Delta ^2+\Omega ^2\right)}
    .
\end{align}
In the tuned case, $S_Q^\text{(wave)}$ is saturated by $S_{h,h}$ in Eq.~\ref{eq:TunedPSD}. The intuition behind the QCRB in Eq.~\ref{eq:wfQCRB} is that a greater uncertainty in $\hat x^\text{(cav)}$ allows for a smaller uncertainty in $\hat p^\text{(cav)}$ by the Heisenberg Uncertainty Principle. This provides a smaller error in the estimation of $\tilde h$. In the detuned case, one intuition behind why problems with incompatibility arise is that the probe observable itself is driven by the signal. This occurs because the detuning mixes the quadratures. Ref.~\cite{Miao+2017} finds a condition $\im{[\hat F(\Omega), \hat F^\dag(\Omega)]}\neq0$ for when the waveform QCRB is loose which holds here for the detuned interferometer. The gap at the detuning frequency $\Omega = \Delta$ is
\begin{align}\label{eq:TheGapAtDetFreq}
    \frac{S^\text{standard}_{h,h}}{S_Q^\text{(wave)}} &= \frac{2}{1+G} \in (1, 2) \\
    \intertext{where}
    S^\text{standard}_{h,h} &= \min_{\theta\in[0, 2\pi)} \frac{1}{\pi T} \var{\frac{\hat x_\theta(\Omega)}{d_\theta}} \\
    \intertext{and}
    G &= \frac{\gamma\sqrt{\gamma^2+4\Delta^2}}{\gamma^2+2\Delta^2}\in(0,1)
    .
\end{align}
Here, $S^\text{standard}_{h,h}$ is the sensitivity $S_{h,h}$ from the standard variational readout scheme where the readout angle $\theta$ is optimised for each $\Omega$. Ref.~\cite{Miao+2017} understood the gap as ultimately arising from the Generalised Schr\"odinger-Heisenberg Uncertainty Principle but did not find a tight fundamental bound or determine the optimal measurement scheme. Our formalism in the Letter addresses the gap by focusing on simultaneous and weighted estimation of $A$ and $B$ instead of complex estimation of $\tilde h$.

\section{Analogy to a toy model}
This section accompanies the section titled \emph{A toy model} in the Letter.

\subsection{Displacements for the detuned interferometer}
Expanding the signal term in Eq.~\ref{eq:ComplexIOVectors} into real and imaginary parts using Eq.~1 provides that
\begin{align}
    \vec{d_h} \tilde h &= 
    \left( \re{\vec{d_h}} + i \im{\vec{d_h}} \right) \pi T (A + i B) \\
    &=    \pi T \left( \re{\vec{d_h}} + i \im{\vec{d_h}} \right) A
    \\ &+ \pi T \left( -\im{\vec{d_h}}+ i \re{\vec{d_h}} \right) B\nonumber
    .
\end{align}

This determines that the signal displacements in Eq.~4 are

\begin{align}\label{eq:dAanddB}
    \vec{d}_A
    &= \sqrt{\pi T} \bmatrixByJames{ \re{\vec{d_h}} \\ \im{\vec{d_h}} } \\
    &= \frac{\sqrt{2\gamma\pi T} g}{\hbar  \left(\gamma ^2+(\Delta -\Omega )^2\right) \left(\gamma ^2+(\Delta +\Omega )^2\right)}
    \\&\cdot 
\bmatrixByJames{
 \Delta  \left(\gamma ^2+\Delta ^2-\Omega ^2\right) \\
 -\gamma  \left(\gamma ^2+\Delta ^2+\Omega ^2\right) \\
 2 \gamma  \Delta  \Omega  \\
 -\Omega  \left(\gamma ^2-\Delta ^2+\Omega ^2\right) \\
} \nonumber\\
\intertext{and}
    \vec{d}_B &= \sqrt{\pi T} \bmatrixByJames{-\im{\vec{d_h}} \\ \re{\vec{d_h}}} \\
    &= \frac{\sqrt{2\gamma\pi T} g}{\hbar  \left(\gamma ^2+(\Delta -\Omega )^2\right) \left(\gamma ^2+(\Delta +\Omega )^2\right)}
    \\&\cdot
\bmatrixByJames{
 -2 \gamma  \Delta  \Omega  \\
 \Omega  \left(\gamma ^2-\Delta ^2+\Omega ^2\right) \\
 \Delta  \left(\gamma ^2+\Delta ^2-\Omega ^2\right) \\
 -\gamma  \left(\gamma ^2+\Delta ^2+\Omega ^2\right) \\
}\nonumber
    .
\end{align}
The factor of $\sqrt{\pi T}$ above arises from the factor of $\frac{1}{\sqrt{\pi T}}$ in Eq.~4 and the factor of $\pi T$ in Eq.~1. The signal displacements each have support in all four components, are orthogonal, and have common Euclidean norm
\begin{align}
    \eta &= \norm{\vec{d}_A} \\&= \norm{\vec{d}_B} \\
    &= \sqrt{\pi T} \sqrt{\norm{\re{\vec{d_h}}}^2 + \norm{\im{\vec{d_h}}}^2} \\
    &= \frac{\sqrt{2\gamma \pi T} g}{\hbar } \sqrt{\frac{  \left(\gamma ^2+\Delta ^2+\Omega ^2\right)}{\left(\gamma ^2+(\Delta -\Omega )^2\right) \left(\gamma ^2+(\Delta +\Omega )^2\right)}}
    .
\end{align}
The commutator of $\h A_\text{only}$ and $\h B_\text{only}$, and therefore $\mu$, can also be found from the above expressions for the displacements.

\subsection{Noise in the output light}
The noise in Eq.~4 is 
\begin{align}
    &\vec{\hat q}^{\,(0)} = \text{M}_{q^\text{(in)}}^{q^\text{(0)}} \vec{\hat q}^{\,\text{(in)}}
\end{align}
Here, $\vec{\hat q}^{\,\text{(in)}}$ is defined similarly to $\vec{\hat q}$ in Eq.~4 from, e.g., $\hat x^\text{(0)}(\Omega)$. $\text{M}_{q^\text{(in)}}^{q^\text{(0)}}$ is derived from $\text{M}_\text{(in)}^\text{(0)}$ in Eq.~\ref{eq:Min0} and is given by
\begin{widetext}
\begin{align}
    \text{M}_{q^\text{(in)}}^{q^\text{(0)}} &= \frac{1}{\left(\gamma ^2+(\Delta -\Omega )^2\right) \left(\gamma ^2+(\Delta +\Omega )^2\right)}\\
    &\cdot
    \bmatrixByJames{
 \gamma ^4-\left(\Delta ^2-\Omega ^2\right)^2 & -2 \gamma  \Delta  \left(\gamma ^2+\Delta ^2-\Omega ^2\right) & -2 \gamma  \Omega  \left(\gamma ^2-\Delta ^2+\Omega ^2\right) & 4 \gamma ^2 \Delta  \Omega  \\
 2 \gamma  \Delta  \left(\gamma ^2+\Delta ^2-\Omega ^2\right) & \gamma ^4-\left(\Delta ^2-\Omega ^2\right)^2 & -4 \gamma ^2 \Delta  \Omega  & -2 \gamma  \Omega  \left(\gamma ^2-\Delta ^2+\Omega ^2\right) \\
 2 \gamma  \Omega  \left(\gamma ^2-\Delta ^2+\Omega ^2\right) & -4 \gamma ^2 \Delta  \Omega  & \gamma ^4-\left(\Delta ^2-\Omega ^2\right)^2 & -2 \gamma  \Delta  \left(\gamma ^2+\Delta ^2-\Omega ^2\right) \\
 4 \gamma ^2 \Delta  \Omega  & 2 \gamma  \Omega  \left(\gamma ^2-\Delta ^2+\Omega ^2\right) & 2 \gamma  \Delta  \left(\gamma ^2+\Delta ^2-\Omega ^2\right) & \gamma ^4-\left(\Delta ^2-\Omega ^2\right)^2 \\}
.\nonumber
\end{align}
\end{widetext}
The covariance matrix $\cov{\vec{\hat q}^{\,(0)}}{\vec{\hat q}^{\,(0)}}$ is the same as the ground state of two harmonic oscillators since, e.g.,

\begin{align}
\label{eq:variancesEqual}
    \var{{\vec{\hat q}\,}_1} &= \var{\hat X_1} = \frac{1}{2}\\
    \intertext{where the related variances for the outgoing field are} 
    \var{\hat x(t')} &= \frac{1}{2}\delta_t(0) \\ 
    \var{\hat x(\Omega)} &= \pi T \\ 
    \var{\re{\hat x(\Omega)}} &= \frac{1}{2}\pi T 
    .
\end{align}

The rest of the diagonal components of the respective covariance matrices are equal and all off-diagonal terms are zero.

\subsection{Converting between the different QCRBs}
For the detuned interferometer, the QCRB for estimation of $A$ and $B$ is weight-invariant because the displacements' norm $\eta$ is common. The QCRB is

\begin{align}\label{eq:QCRBDetIFO} 
    \Sigma_Q &= \frac{1}{\eta^2} \\
    &= \frac{1}{\pi T}\frac{\hbar^2}{2 g^2}
    \frac{\left(\gamma ^2+(\Delta -\Omega )^2\right) \left(\gamma ^2+(\Delta +\Omega )^2\right) }{\gamma  \left(\gamma ^2+\Delta ^2+\Omega ^2\right)}\\
    &= \frac{1}{\pi T} S_Q^\text{(wave)}
    .
\end{align}
Here, $S_Q^\text{(wave)}$ from Eq.~\ref{eq:wfQCRBwithLittleG} is the QCRB for the estimation of the complex signal $\tilde{h}$. Compare this to the spectral form of the QCRB for simultaneous estimation of $A$ and $B$
\begin{align}\label{eq:QCRBDetIFOspectral}
    S_Q &= \frac{1}{\pi T}\Sigma_Q \\
    &= \frac{1}{\pi^2 T^2}\frac{\hbar^2}{2 g^2}
    \frac{\left(\gamma ^2+(\Delta -\Omega )^2\right) \left(\gamma ^2+(\Delta +\Omega )^2\right) }{\gamma  \left(\gamma ^2+\Delta ^2+\Omega ^2\right)}\\
    &= \frac{1}{\pi^2 T^2} S_Q^\text{(wave)}
    .
\end{align}
The factors of $\pi T$ appear above because the longer the integration time of a sinusoidal signal at frequency $\Omega$, the larger the absolute value of the Fourier component at $\Omega$ becomes by Eq.~1.

Since the benefits of a longer integration time are already well-understood, in Fig.~2, we use effective amplitude spectral density units without factors of integration time. These units are consistent with $\sqrt{S_Q^\text{(wave)}}$ and the strain sensitivity of the standard variational readout scheme. We define the comparable estimation error and HCRB in power units as
\begin{align}\label{eq:EffEffQuantities}
    S^\text{(wave)}   &= \pi^2 T^2 S   = \pi T \Sigma   \\
    S_H^\text{(wave)} &= \pi^2 T^2 S_H = \pi T \Sigma_H
    .
\end{align}
These can be interpreted as spectral quantities defined with respect to the frequency-domain real and imaginary parts instead of $A$ and $B$. We omit this subtlety of further removing factors of the integration time from the Letter for clarity. 
Ultimately, the ratio between the quantities, e.g.\ the gap from the HCRB to the QCRB, is more important and is invariant under such a change of units.

\subsection{Transformation to the toy model}
In Eq.~10, the linear transformation to the toy model is defined by a symplectic 4-by-4 matrix $\text{M}$ that preserves the commutator structure and the covariance matrix. Although $\hat X^{(0)}\neq\hat q^{(0)}$, the noise's mean and covariance matrix are preserved which is sufficient for Gaussian estimation. The transformation $\text{M}$ can be found uniquely by requiring that the normalised displacements (e.g.\ $\vec{n}_A$) and their conjugates (e.g.\ $\vec{n}_A^*$) are respected such that
\begin{align}
    \vec{n}_A = \text{M}\frac{\vec{d}_A}{\eta}&, \quad
    \vec{n}_B  = \text{M}\frac{\vec{d}_B}{\eta} \\
    \intertext{and the conjugates are}
    \vec{n}_A^* = \text{M}\frac{\vec{d}_A^*}{\eta} = \bmatrixByJames{ 0 \\ 1 \\ 0 \\ 0} &,\quad
    \vec{n}_B^*  = \text{M}\frac{\vec{d}_B^*}{\eta} = \begin{bmatrix} -\mu \\ 0 \\ 0 \\ \sqrt{1-\mu^2}\end{bmatrix}.
\end{align}
Here, the conjugate vectors are defined by mapping canonically conjugate variables to each other, e.g.\ $\hat X_j\mapsto\hat P_j$ and $\hat P_j\mapsto-\hat X_j$ for $j=1,2$.

At the detuning frequency, the transformation is

\begin{widetext}
\begin{align}\label{eq:MAtDetFreq}
    \text{M} = \frac{1}{\sqrt{\gamma ^4+6 \gamma ^2 \Delta ^2+8 \Delta ^4}}
   \bmatrixByJames{
 -2 \Delta ^2 & \gamma  \Delta  & \gamma  \Delta  & -\gamma ^2-2 \Delta ^2 \\
 -\gamma  \Delta  & -2 \Delta ^2 & \gamma ^2+2 \Delta ^2 & \gamma  \Delta  \\
 \Delta  \sqrt{\gamma ^2+4 \Delta ^2} & -\gamma  \sqrt{\gamma ^2+4 \Delta ^2} & 0 & -\Delta  \sqrt{\gamma ^2+4 \Delta ^2} \\
 \gamma  \sqrt{\gamma ^2+4 \Delta ^2} & \Delta  \sqrt{\gamma ^2+4 \Delta ^2} & \Delta  \sqrt{\gamma ^2+4 \Delta ^2} & 0 \\}
    .
\end{align}    
\end{widetext}
This matrix is orthogonal such that $\text{M}^{-1} = \text{M}^\text{T}$. At frequencies $\Om\neq\Delta$, we find the transformation numerically.

\subsection{Displacements for the toy model}

\begin{figure}[ht]
    \includegraphics[width=\columnwidth]{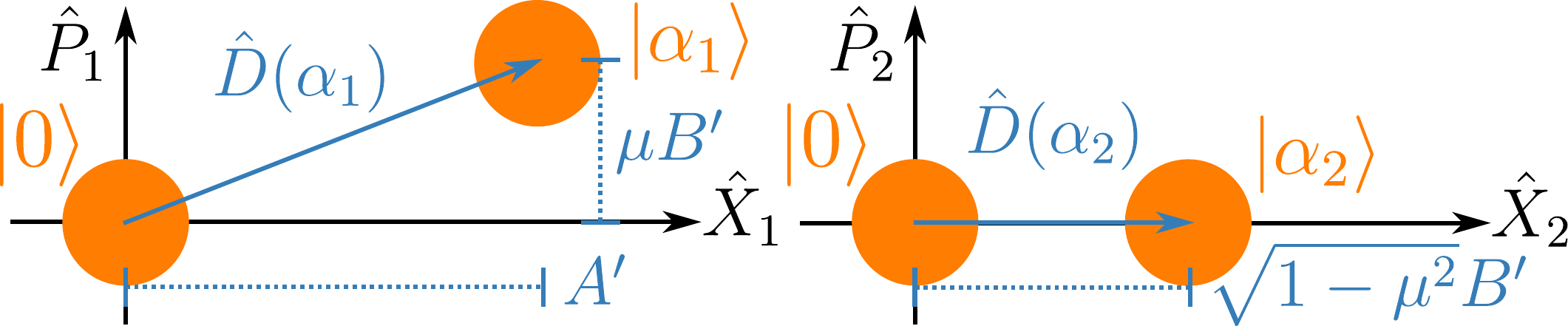}
    \caption{Phase space representation of the toy model of two displaced harmonic oscillators, analogous to the detuned interferometer at each positive frequency.}
    \label{fig:S1}
\end{figure}

In Eq.~8, the incompatible displacements for the toy model with respect to $A'$ and $B'$, respectively, are
\begin{align}
    \vec{n}_A\cdot\vec{\hat X} &= \hat X_1, \quad
    \vec{n}_B\cdot\vec{\hat X} = \mu\hat P_1 + \sqrt{1-\mu^2}\hat X_2
    .
\end{align}
These should not be confused with those acting on each oscillator $\hat D(\alpha_1)$ and $\hat D(\alpha_2)$ which commute and are shown in Fig.~\ref{fig:S1}. The final state is a pure separable two-mode coherent state with the density operator
\begin{align}\label{eq:coherentAmplitudes}
    \hat\rho &= \ket{\psi}\bra{\psi} \\
    \where
    \ket{\psi} &= \ket{\alpha_1}\otimes\ket{\alpha_2} \\
    &= \hat D(\alpha_1)\ket{0}\otimes\hat D(\alpha_2)\ket{0} \\
    \intertext{and}
    \hat D(\alpha) &= \exp\left(\alpha \hat a^\dag  - \alpha^\ast \hat a \right)\\   
    \alpha_1 &= \frac{1}{\sqrt2}(A'+i\mu B'), \quad \alpha_2 = \frac{1}{\sqrt2}\sqrt{1-\mu^2}B'\\
    \suchthat
    \hat D(\alpha_1) &= \exp\left(-i A' \hat P_1 + i \mu B' \hat X_1 \right) \\    
    \hat D(\alpha_2) &= \exp\left(-i \sqrt{1-\mu^2}B' \hat P_2 \right)
    .
\end{align}
Here, $\ket{0}$ is the vacuum state of a harmonic oscillator and we omit a normalisation factor for the coherent states by assuming that $A'$ and $B'$ are small. The commutator of the displacements in Eq.~8 has the opposite sign to the detuned interferometer. We can either ignore the sign because the absolute value of the commutator is what matters, or we can implicitly exchange $A'\leftrightarrow B'$ to preserve the sign. We prefer the latter approach, but the results are the same either way.

\section{The Holevo Cram\'er-Rao Bound (HCRB)}
This section accompanies the section titled \emph{Results for the toy model} in the Letter.

\subsection{Introduction and definition of the HCRB}

Given a weight matrix $\text{W}$ and state $\hat\rho$ that depends on parameters
$\{ A'_{j} \}_{j}$,
the Holevo Cram\'er-Rao bound (HCRB) sets a lower bound on the weighted simultaneous estimation error of the parameters. It is defined as~\cite{Holevo2011book} 
\begin{align}\label{eq:HCRB}
\Sigma_H' &=
\min_{\vec{\hat Y}=(\hat Y_1, \hat Y_2)^\text{T}\in\mathcal{Y}}\left(\Sigma_{H,\text{obj}}'\right) \\
\intertext{where the objective function is}
\Sigma_{H,\text{obj}}' &=
\tr{\text{W}\re{\text{M}_Y}}+ \trabs{\sqrt{\text{W}}\im{\text{M}_Y}\sqrt{\text{W}}} 
\intertext{and $\text{M}_Y$ is the covariance matrix of $\vec{Y}$ with respect to $\hat{\rho}$}
    \text{M}_Y &=
    \bmatrixByJames{
\tr{\h\rho\hat Y_1^2} & \tr{\h\rho\hat Y_1\hat Y_2} \\[5pt] \tr{\h\rho\hat Y_2\hat Y_1} & \tr{\h\rho\hat Y_2^2}}
\end{align}
where the observables $\vec{Y}$ must obey the following constraints
\begin{align}
\label{eq:Yconstraints}
    \mathcal{Y} &=
        \left\{
            \vec{\hat Y}:\; \hat Y_j^\dag = \hat Y_j,\; \tr{\h\rho\hat Y_j}=0,\; \tr{\hat Y_k\partial_{\vec{A'}_j}\hat\rho} = \delta_{j,k}
        \right\}.
\end{align}

Here, $\text{TrAbs}$ of an operator is defined as the sum of the absolute values of its eigenvalues and the signals are $\vec{A'}=\left(A', B'\right)^\text{T}$.

The constraints on $\left(\hat{Y}_{1},\hat{Y}_{2}\right)$ mean that they are unbiased error observables, e.g.\ $\hat{Y}_{1}$ takes the form of $\hat A' - A'\hat 1$, where $\hat A'$ is a locally unbiased estimate of $A'$. 

Solving Eq.~\ref{eq:HCRB} for Gaussian displacements is a semi-definite program a priori~\cite{Bradshaw+2018,doi:10.1137/1038003}. Here, it would involve 4-by-4 real matrices. \\

The hierarchy of Cram\'er-Rao bounds is
\begin{align}\label{eq:hierarchy}
    \Sigma'\geq\Sigma_C'\geq\Sigma_H'\geq\Sigma_Q'.
\end{align}
Here, the weighted CCRB $\Sigma_C'$ is defined by the classical Fisher information matrix $\text{U}$ as~\cite{WisemanMilburn2009book}
\begin{align}\label{eq:wCCRB}
    \Sigma_C'&=\tr{\text{W} \text{U}^{-1}}\\
    \intertext{where}
    \text{U}_{j,k}
    &=
    \sum_{l=1}^2
    \frac{
    \abs{\partial_{\vec{A}'_j}\ev{\vec{\hat A}'_l}}\abs{\partial_{\vec{A}'_k}\ev{\vec{\hat A}'_l}}
    }{
    \var{\vec{\hat A}'_l}
    }
    \intertext{and the unbiased estimates are}
    \vec{\h A}' &= \bmatrixByJames{\h A' \\ \h B'}
    .
\end{align}
The HCRB is saturated for non-commuting Gaussian displacements by the optimal Gaussian estimates~\cite{Holevo2011book} such that Eq.~\ref{eq:hierarchy} becomes
\begin{align}\label{eq:GaussianHierarchy}
    \Sigma'=\Sigma_C'=\Sigma_H'>\Sigma_Q'.
\end{align}

\subsection{Semi-analytic result for the HCRB}

For the toy model, we calculate the HCRB using the procedure from Ref.~\cite{Bradshaw+2017}. Other approaches to solving the semi-definite program derived from Eq.~\ref{eq:HCRB} are also viable. E.g., taking the limit of zero thermal occupation of the procedure in Ref.~\cite{Bradshaw+2018}. We confirm our results with an independent numerical optimisation of the CCRB in Eq.~\ref{eq:wCCRB} over arbitrary compatible linear combinations of $\vec{\hat X}$ which suffices given Eq.~\ref{eq:GaussianHierarchy}. \\ 

We can assume an operating point of zero around which the signals $A'$ and $B'$ are small without loss of generality. This is justified by a similar argument to that given in Ref.~\cite{Bradshaw+2017}. The HCRB is an asymptotic bound in the number of observations $N$ such that a non-zero operating point can be adaptively estimated by the first $\sqrt{N}$ observations and then subtracted to leave the signals small for the last $N - \sqrt{N}$ observations. This allows for a linear approximation to the final state in Eq.~\ref{eq:coherentAmplitudes} as
\begin{align}
    \ket{\psi} &\approx \ket{\psi_0} + A'\ket{\psi_1} + B'\ket{\psi_2} \\
    \where
    \ket{\psi_0} &= \ket{0}\otimes\ket{0} \\
    \ket{\psi_1} &= -i \hat P_1 \ket{0} \otimes \ket{0}\\
    \ket{\psi_2} &= i \left(\mu \hat X_1 - \sqrt{1-\mu^2} \hat P_2\right) \ket{0} \otimes \ket{0} \\
    \suchthat
    \hat\rho &= \ket{\psi}\bra{\psi} \\ &\approx \ket{\psi_0}\bra{\psi_0} + 2A'\re{\ket{\psi_1}\bra{\psi_0}} \\&+ 2B'\re{\ket{\psi_2}\bra{\psi_0}}\nonumber
    .
\end{align}
Here, $\ket{\psi_j}$ for $j=1,2$ are unnormalised states. The QCRB can be calculated to be uniformly unity from the symmetric logarithmic derivatives~\cite{WisemanMilburn2009book}. The left and right logarithmic derivatives are zero.\\

By using the Gram-Schmidt process, we find an orthonormal basis $\braket{e_k}{e_j}=\delta_{j,k}$ of the $\{\ket{\psi_j}\}_j$ subspace.
\begin{align}\label{eq:basisCoeffs}
    \ket{e_0} &= \ket{\psi_0} \\
    \ket{e_1} &= \sqrt{2} \ket{\psi_1} \\
    \ket{e_2} &= \sqrt{\frac{2}{1 - \mu^2}}(\ket{\psi_2} - i\mu \ket{\psi_1})
    .
\end{align}
In the limit of $\mu=1$, $\ket{e_2}=0$ since the $\{\ket{\psi_j}\}_j$ subspace is only two-dimensional. Without loss of generality, this subspace contains the support of $\hat Y_1$ and $\hat Y_2$ in the minimisation in Eq.~\ref{eq:HCRB}. The constraints of $\mathcal{Y}$ in Eq.~\ref{eq:Yconstraints} can be expanded into this orthonormal basis as
\begin{align}
    \braopket{e_0}{\hat Y_j}{e_0} &= 0, \quad j=1,2 \\
    2\re{\braopket{\psi_k}{\hat Y_j}{e_0}} &= \delta_{j,k}, \quad j,k\in\{1,2\}\label{eq:2ndConstraintIndexNotation}
    .
\end{align}
Let $y_{j,k} = \braopket{e_0}{\hat Y_j}{e_k}$ be complex numbers. The constraint in Eq.~\ref{eq:2ndConstraintIndexNotation} is then
\begin{align}
    \label{eq:2ndConstraint}
    \bmatrixByJames{\frac{1}{\sqrt2} & 0 \\ 0 & \frac{1}{\sqrt2}} &=
    \bmatrixByJames{
    \re{y_{1,1}} &
    \sqrt{1 - \mu^2}\re{y_{1,2}} - \mu \im{y_{1,1}} \\
    \re{y_{2,1}} &
    \sqrt{1 - \mu^2}\re{y_{2,2}} - \mu \im{y_{2,1}} \\
    } 
    .
\end{align}

All other independent components of the Hermitian $\hat Y_j$ in this basis are zero without loss of generality. In the $\{\ket{e_j}\}_j$ basis, 
\begin{align}\label{eq:Yests}
    \hat Y_1 &= \bmatrixByJames{
    0 & y_{1,1} & y_{1,2} \\
    y_{1,1}^* & 0 & 0 \\
    y_{1,2}^* & 0 & 0 \\    
    } \\
    \hat Y_2 &= \bmatrixByJames{
    0 & y_{2,1} & y_{2,2} \\
    y_{2,1}^* & 0 & 0 \\
    y_{2,2}^* & 0 & 0 \\    
    }
    .
\end{align}
The objective of the minimisation in Eq.~\ref{eq:HCRB} is
\begin{align}\label{eq:ObjFnUnconstrained}
    \Sigma_{H,\text{obj}}' &= 2w\left(\abs{y_{1,1}}^2 + \abs{y_{1,2}}^2\right)\\
    &+ 2(1 - w)\left(\abs{y_{2,1}}^2 + \abs{y_{2,2}}^2\right)\nonumber\\
    &+ 4\sqrt{w(1 - w)}\abs{z}\nonumber\\
    \intertext{where}
    z &= y_{2,1}y_{1,1}^* + y_{2,2}y_{1,2}^*\\
    &= \re{y_{1,1}} \im{y_{2,1}} - \im{y_{1,1}} \re{y_{2,1}} \\&+ \re{y_{1,2}} \im{y_{2,2}} - \im{y_{1,2}} \re{y_{2,2}} \nonumber
    .
\end{align}
By Eq.~\ref{eq:2ndConstraint}, $\re{y_{1,1}} = \frac{1}{\sqrt2}$ and $\re{y_{2,1}}$ is zero. We observe that the minimum is achieved with $z=0$ without loss of generality. Intuitively, the $\abs{z}$ term in the objective corresponds to the penalty for incompatibility. 
$z=0$ can be achieved with $\re{y_{1,2}} = \im{y_{1,1}} = 0$ and the following substitutions
\begin{align}
    \re{y_{2,2}} &= \kappa \cos(\phi), \quad
    \im{y_{2,1}} = \kappa \sin(\phi) \\ 
    \im{y_{1,2}} &= \frac{\re{y_{1,1}} \im{y_{2,1}}}{\re{y_{2,2}}} = \frac{1}{\sqrt2} \tan(\phi)
    \intertext{subject to the remaining constraint from Eq.~\ref{eq:2ndConstraint}}
     \frac{1}{\sqrt2}  &= \kappa\cos(\nu+\phi)\\ 
    \intertext{where}
    \kappa &= \sqrt{\re{y_{2,2}}^2 + \im{y_{2,1}}^2} \\
    \phi &= \arctan\left(\frac{\im{y_{2,1}}}{\re{y_{2,2}}}\right) \\
    \nu &= \arcsin(\mu) \label{eq:NuDef}
    .
\end{align}
Here, $\phi\in(0,\pi]$ and $\nu\in(0,\frac{\pi}{2}]$.
The objective function in Eq.~\ref{eq:ObjFnUnconstrained} then becomes
\begin{align}
     \Sigma_{H,\text{obj}}' &= 
     2 w \left(\frac{1}{2} + \frac{1}{2}\tan(\phi)^2\right)
     +2 (1-w) \frac{1}{2\cos(\nu+\phi)^2}\\
     &= 
     w \frac{1}{\cos(\phi)^2}
     + (1-w) \frac{1}{\cos(\nu+\phi)^2}
    .  \label{eq:ObjFnSemiAnalytic}
\end{align}
This is the single-parameter minimisation in Eq.~11.

\subsection{Analytic results for the HCRB}

Previous work has solved Eq.~\ref{eq:HCRB} in the limit of $\mu=1$ and $w=\frac{1}{2}$ (although without using this terminology)~\cite{Bradshaw+2017}. We derive the analytic results in Eq.~12 in the limit of $\mu=1$ and arbitrary $w$, and in the limit of $w=\frac{1}{2}$ and arbitrary $\mu$.\\

\emph{Limit of $\mu=1$ and arbitrary $w$.} --- In this limit, the objective function in Eq.~\ref{eq:ObjFnSemiAnalytic} becomes
\begin{align}
\Sigma_{H,\text{obj}}' = \frac{w}{\cos(\phi)^{2}}+\frac{1-w}{\sin(\phi)^{2}}.
\end{align}
By the Cauchy-Schwarz inequality, we can find a lower bound on the objective function
\begin{align}
&1 \cdot \left(\frac{w}{\cos(\phi)^{2}}+\frac{1-w}{\sin(\phi)^{2}}\right) \\
&\geq\left(\cos\left(\phi\right)\frac{\sqrt{w}}{\cos\left(\phi\right)}+\sin\left(\phi\right)\frac{\sqrt{1-w}}{\sin\left(\phi\right)}\right)^{2} \nonumber\\
&= \left(\sqrt{w}+\sqrt{1-w}\right)^{2}\\
&= 1+2\sqrt{w(1-w)}
.
\end{align}
Since the Cauchy-Schwarz inequality is tight, we know that this is the optimal value. This provides the HCRB in Eq.~12 in this limit. It is saturated by the optimal angle $\barBetter{\phi}$ satisfying

\begin{align}
\frac{\sqrt{w}}{\cos\left(\barBetter{\phi}\right)^{2}}&=\frac{\sqrt{1-w}}{\sin\left(\barBetter{\phi}\right)^{2}}
\end{align}
such that
\begin{align}
\barBetter{\phi} &= \phi_0 = \arctan\left(\left(\frac{1-w}{w}\right)^{\frac{1}{4}}\right)
\label{eq:phi_opt_mu_1}
\intertext{or}
\barBetter{\phi} &= \pi- \phi_0
.
\end{align}

\emph{Limit of $w=\frac{1}{2}$ and arbitrary $\mu$.} --- In this limit, the objective function in Eq.~\ref{eq:ObjFnSemiAnalytic} becomes
\begin{align}
\Sigma_{H,\text{obj}}' = 
\frac{1}{2\cos\left(\phi\right)^{2}}+\frac{1}{2\cos\left(\phi+\nu\right)^{2}}.    
\end{align}
Again, by the Cauchy-Schwarz inequality,

\begin{align}
&\left(
2\cos\left(\phi\right)^2
+
2\cos\left(\phi + \nu\right)^2
\right)
\left(
\frac{1}{2\cos\left(\phi\right)^2}
+
\frac{1}{2\cos\left(\phi + \nu\right)^2}
\right)
\\
&\geq
\left(
    \sqrt{2}\cos\left(\phi\right)\frac{1}{\sqrt{2}\cos\left(\phi\right)}
    +
    \sqrt{2}\cos\left(\phi + \nu\right)\frac{1}{\sqrt{2}\cos\left(\phi + \nu\right)}
\right)^{2}\label{eq:CSBineqUse1}
\\
&= 4
.
\end{align}
such that
\begin{align}
\Sigma_{H,\text{obj}}'
&\geq
\frac{2}{\cos\left(\phi\right)^{2}+\cos\left(\phi+\nu\right)^{2}}\\
&\geq\frac{1}{\cos\left(\frac{\nu}{2}\right)^{2}}\label{eq:ATightBound}
\intertext{where we used the fact that, since $\nu\in(0,\frac{\pi}{2}]$, }
2\cos\left(\frac{\nu}{2}\right)^{2}
&\geq
\cos\left(\phi\right)^{2}+\cos\left(\phi+\nu\right)^{2}
\label{eq:TrigTrickWithNu}
.
\end{align}
Since the Cauchy-Schwarz inequality in Eq.~\ref{eq:CSBineqUse1} and the trigonometric inequality in Eq.~\ref{eq:TrigTrickWithNu} are both saturated for $\barBetter{\phi}=\pi-\frac{\nu}{2}$, the bound on the objective function in Eq.~\ref{eq:ATightBound} is tight. Furthermore, by Eq.~\ref{eq:NuDef},
\begin{align}
    \cos\left(\frac{\nu}{2}\right)^{2}=\frac{1}{2}\left(1+\sqrt{1-\mu^{2}}\right)
    \intertext{such that the HCRB in Eq.~12 in this limit is}
    \Sigma_H' = \frac{2}{1+\sqrt{1-\mu^{2}}}.
\end{align}
We summarize the analytic expressions for the HCRB and $\barBetter{\phi}$ in Table~\ref{tab:analytics}.\\

\setlength{\tabcolsep}{0.5em}
\begin{table}[h]
    \centering
    \begin{tabular}{@{}cc|cc@{}}
    \toprule
    $\mu$ & $w$ & HCRB, $\Sigma_H'$ & optimal angle $\barBetter{\phi}$ \\
    \hline \\ [-2ex]
    $1$ & any  & $1+2\sqrt{w} \sqrt{1-w}$  & $\arctan\left(\left(\frac{1-w}{w}\right)^{\frac{1}{4}}\right)$  \\[5pt]
    any & $\frac{1}{2}$  & $\frac{2}{1+\sqrt{1-\mu^{2}}}$  & $\pi-\frac{\nu}{2}$  \\
    \bottomrule
    \end{tabular}
    \caption{Analytic solutions for the HCRB and optimal angle $\barBetter{\phi}$ in certain limits.}
    \label{tab:analytics}
\end{table}

The general solution for the optimal angle $\barBetter{\phi}$ in Eq.~11 for $\mu > 0$ reduces to finding the roots of the following order-8 polynomial in $t = \tan\left(\frac{\barBetter{\phi}}{2}\right)$
\begin{align}\label{eq:semianalyticSolOctivePolynomial}
    p(t)&=t^8 ((-2+2w) \sin (\nu))\\
    &+t^7 ((4-w) \cos (\nu)+w \cos (3\nu))\nonumber\\
    &+t^6 ((8-2 w) \sin (\nu)+6 w \sin (3\nu))\nonumber\\
    &+t^5 ((-12+15w) \cos (\nu)-15 w \cos (3\nu))\nonumber\\
    &+t^4 ((-12+24 w) \sin (\nu)-20 w \sin (3\nu))\nonumber\\
    &+t^3 ((12-15 w) \cos (\nu)+15 w \cos (3\nu))\nonumber\\
    &+t^2 ((8-2 w) \sin (\nu)+6 w \sin (3\nu))\nonumber\\
    &+t ((-4+w ) \cos (\nu)-w \cos (3\nu))\nonumber\\
    &+(-2+2 w) \sin (\nu)\nonumber
    .
\end{align}
We found this via brute force using \textsc{Mathematica}~\cite{mathematica}.
Since there is no closed solution for such polynomials, by the Abel–Ruffini Theorem, we do not have a general closed solution for the HCRB.\\

\begin{figure}[t]
    \includegraphics[width=\columnwidth]{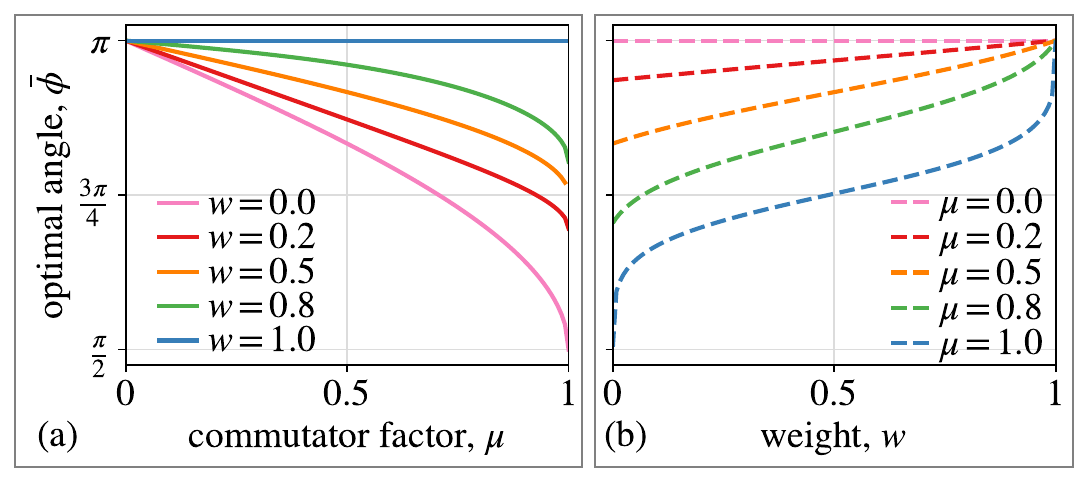}
    \caption{Optimal angle $\barBetter{\phi}$ in Eq.~11 versus (a) the commutator $\mu$ for different weights $w$ and (b) the weight $w$ for different commutators $\mu$.}
    \label{fig:S2}
\end{figure}

The optimal angle $\barBetter{\phi}$ found numerically is shown in Fig.~\hyperref[fig:S2]{S2a} versus $\mu$ for different weights $w$. For the single-parameter estimation limit of $w=0$ ($w=1$), the optimal angle $\barBetter{\phi}$ is $\pi-\nu$ ($\pi$). For a given value of $\mu$, as $w$ increases from $0$ to $1$, $\barBetter{\phi}$ increases monotonically from $\pi-\nu$ to $\pi$. This is shown in Fig.~\hyperref[fig:S2]{S2b} which shows $\barBetter{\phi}$ versus $w$ for different $\mu$. For $\mu \ll 1$, $\barBetter{\phi}$ drops linearly as $\barBetter{\phi} \approx \pi-\nu\left(1-w\right)$. For larger $\mu$, the $w$ dependence shown in Fig.~\hyperref[fig:S1]{S1b} is nonlinear and, in the limit of $\mu=1$, satisfies Eq.~\ref{eq:phi_opt_mu_1}.

\subsection{Optimal estimates for the toy model}

We derive a closed form for the optimal estimates. The CCRB in Eq.~\ref{eq:wCCRB} is degenerate within the Quantum Mechanics--free subspace of the optimal estimates. We first find the optimal subspace ($\hat T_1$ and $\hat T_2$ in Eq.~13) and then project the displacements onto the subspace to find the optimal unbiased estimates as
\begin{align}
    \bmatrixByJames{\h A' \\ \h B'}
    &= \bmatrixByJames{
        \frac{\partial_{A'}\ev{\hat T_1}}{(\partial_{A'}\ev{\hat T_1})^2+(\partial_{A'}\ev{\hat T_2})^2} & 
        \frac{\partial_{A'}\ev{\hat T_2}}{(\partial_{A'}\ev{\hat T_1})^2+(\partial_{A'}\ev{\hat T_2})^2} \\
        \frac{\partial_{B'}\ev{\hat T_1}}{(\partial_{B'}\ev{\hat T_1})^2+(\partial_{B'}\ev{\hat T_2})^2} & 
        \frac{\partial_{B'}\ev{\hat T_2}}{(\partial_{B'}\ev{\hat T_1})^2+(\partial_{B'}\ev{\hat T_2})^2} \\    
    }
    \bmatrixByJames{\hat T_1 \\ \hat T_2}
    .
\end{align}
Here, there is no cross-coupling ($\partial_{A'}\ev{\hat T_2}=\partial_{B'}\ev{\hat T_1}=0$) such that the measurements just need to be debiased in Eq.~15.
This also means that measuring $\h T_1$ provides no information about $B'$ or $\h T_2$. Since $\h T_1$ and $\h T_2$ commute, they have a shared eigenspace, but the above fact means that the eigenvalues of $\h T_2$ are degenerate for any given eigenvalue of $\h T_1$.
By Eq.~\ref{eq:GaussianHierarchy}, it suffices to show that the HCRB in Eq.~11 is saturated by the CCRB of $\hat T_1$ and $\hat T_2$. The classical Fisher information matrix is
\begin{align}\label{eq:CFIM}
\text{U}
&=
\bmatrixByJames{
 2 \cos(\barBetter{\phi}) ^2 & 0 \\
 0 & 2 \left(\sqrt{1-\mu^2} \cos (\barBetter{\phi})-\mu \sin (\barBetter{\phi})\right)^2
}
\suchthat
    \Sigma_\text{C}' &= \frac{w}{\cos(\barBetter{\phi}) ^2} + \frac{1-w}{\left(\sqrt{1-\mu^2} \cos (\barBetter{\phi})-\mu \sin (\barBetter{\phi})\right)^2}.
\end{align}
By identifying $\cos(\barBetter{\phi}+\nu)$ in the denominator, therefore, this saturates the HCRB since $\barBetter{\phi}$ was chosen as the optimal angle.

\section{The waveform-estimation HCRB}
This section accompanies the section titled \emph{Precision limits of detuned interferometry} in the Letter.\\

The HCRB for the detuned interferometer is
    
\begin{align}
    \Sigma_H
    &=\eta^{-2} \Sigma_H'
    \end{align}
    assuming equal weights $w=\frac{1}{2}$ this is
    \begin{align}
    \Sigma_H
    &= \frac{1}{\eta^2} \frac{2}{1+\sqrt{1-\mu^2}} \nonumber\\
    &= \frac{\hbar^2}{4 g^2 \pi  T}\frac{\zeta}{\gamma  \Delta ^2 \Omega ^2} \left(\gamma ^2+\Delta ^2+\Omega ^2-\sqrt{\zeta}\right)  \nonumber
    \where
    \zeta &= \left(\gamma ^2+(\Delta -\Omega )^2\right) \left(\gamma ^2+(\Delta +\Omega )^2\right)
    .
\end{align}
This coincides with the QCRB in Eq.~\ref{eq:QCRBDetIFO} in the limit of a tuned configuration ($\Delta=0$). For Fig.~2, we removed the factors of the integration time to find the effective strain sensitivity $S^\text{(wave)}$ and HCRB $S_H^\text{(wave)}$ in Eq.~\ref{eq:EffEffQuantities} as previously discussed. \\ 

At the detuning frequency and assuming equal weights, the gap from the HCRB to the QCRB is
\begin{align}
    \frac{S_H^\text{(wave)} }{S_Q^\text{(wave)} } \overset{w=\frac{1}{2}}&{=} \frac{2}{1+\sqrt{1-\mu^2}}\\
    \overset{\Omega=\Delta}&{=} \frac{2}{1+G}
    \intertext{such that by Eq.~\ref{eq:TheGapAtDetFreq}}
    S^\text{standard}_{h,h} &= S_H^\text{(wave)}.
\end{align}
This means that the standard variational readout scheme is optimal at $\Om=\Delta$ for equal weights. The numerical results in Fig.~2 show that this is also true at other frequencies for equal weights. For unequal weights, the gap from the sensitivity using variational readout to the QCRB can be at least partially overcome using the optimal measurement scheme. 

\section{Proposal for an experimental realisation}
This section accompanies the section titled \emph{Proposal for an experimental realisation} in the Letter.

\subsection{Time-domain expansion of the optimal estimates}

Our goal is to find an experimental realization of a measurement of the optimal compatible estimates $\hat{A}$ and $\hat{B}$. 
The linear transformation from the toy model, e.g.\ given in Eq.~\ref{eq:MAtDetFreq} at the detuning frequency, provides that in the frequency domain these estimates are
\begin{align}\label{eq:VectorEst}
    \bmatrixByJames{\h A \\ \h B}
    &= \sqrt{\pi T}\text{C} \vec{\hat q}.
\end{align}
Here, $\text{C}$ is a real 2-by-4 matrix determined by Eq.~\ref{eq:OptEstWrtT}, Eq.~13, and $\vec{\hat X} = \text{M}\vec{\hat q}$. 

To search for an experimental realisation, it is valuable to express these estimates in the time domain.
Expanding $\vec{\hat q}$ into the time domain in Eq.~\ref{eq:VectorEst} using Eq.~4 provides that
\begin{align}
    \bmatrixByJames{\h A \\ \h B}
    &= \intginf{t}
    \text{C} \bmatrixByJames{
    \cos(\Omega t) \hat x(t) \\
    \cos(\Omega t) \hat p(t) \\
    \sin(\Omega t) \hat x(t) \\
    \sin(\Omega t) \hat p(t) \\
    }.\label{eq:TimeDomainExpansionFirstStep}  \\
    \intertext{We can show that these correspond to phase and amplitude modulated quadratures. Simplifying Eq.~\ref{eq:TimeDomainExpansionFirstStep} provides that}
    \bmatrixByJames{\h A \\ \h B}&= \intginf{t}
    \bmatrixByJames{
    \beta_{1,1}(t) & \beta_{1,2}(t)  \\
    \beta_{2,1}(t) & \beta_{2,2}(t) \\
    } 
    \bmatrixByJames{
    \hat x(t) \\
    \hat p(t) \\
    }  \label{eq:OptMeasAsModulatedQuadsBetas} \\
    &= \intginf{t}
    \bmatrixByJames{
    c_A(t) \hat x_{\vartheta_A(t)}(t) \\
    c_B(t) \hat x_{\vartheta_B(t)}(t)
    }
    .
\end{align}

This is Eq.~16. Here, the amplitudes $c_A$ and $c_B$ of the time-dependent quadratures are
\begin{align}
    \vec{c}(t) &= \bmatrixByJames{c_A(t) \\ c_B(t)} \\
    \intertext{where for $j,k =1,2$}
    \vec{c}_j(t) &= \sqrt{\beta_{j,1}(t)^2 + \beta_{j,2}(t)^2}\\
    \beta_{j,k}(t) &= \text{C}_{j,k} \cos(\Omega t) + \text{C}_{j,k+2} \sin(\Omega t)
    .    
\end{align}
And, the phases $\vartheta_A$ and $\vartheta_B$ are
\begin{align}
    \vec{\vartheta}(t) &= \bmatrixByJames{\vartheta_A(t) \\ \vartheta_B(t)}\\
    \intertext{where}
    \vec{\vartheta}_j(t) &= \arctan\left(
    \frac{\beta_{j,2}(t)}{\beta_{j,1}(t)}
    \right)
    .    
\end{align}
The time dependence of these phases is not simply sinusoidal.
We emphasise that these time-dependent quadratures are defined in the Interaction Frame with respect to $\h H_0$ and the free evolution of the bath modes in Eq.~\ref{eq:Ham}. The equivalent amplitude and phase modulation outside the Interaction Frame can be obtained from the expressions above.

\subsection{Review of homodyne readout}

\begin{figure}[t]
    \includegraphics[width=0.5\columnwidth]{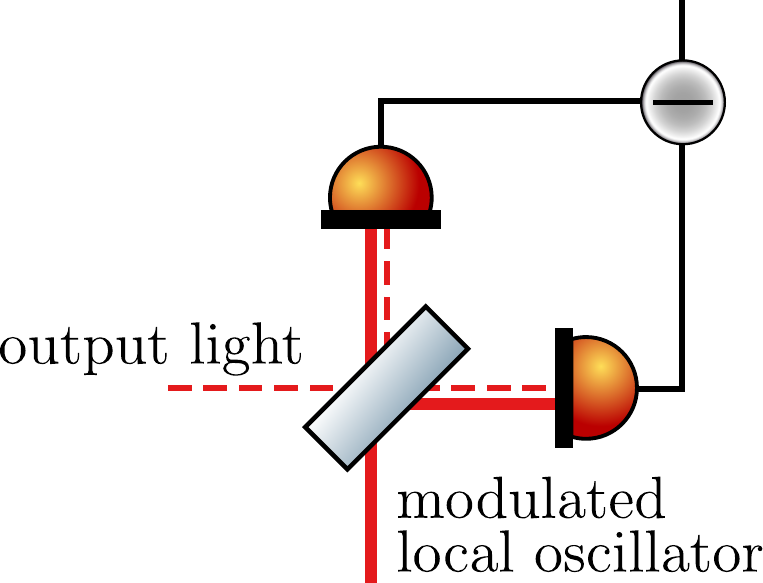}
    \caption{Homodyne readout scheme to measure a quadrature.}
    \label{fig:S3}
\end{figure}

To determine how to measure the modulated quadratures simultaneously, we first need to review how a quadrature measurement can be experimentally realised by homodyne readout.\\

Homodyne readout comprises mixing the beam from the interferometer and a bright local oscillator with coherent amplitude $\alpha(t)$ on a 50:50 beamsplitter, as shown in Fig.~\ref{fig:S3}, then recording the difference current $\h I(t)$ for times $t$ from photodetectors on each of the output beams from the beamsplitter. For a sufficiently bright local oscillator, $\abs{\alpha}\gg1$, then~\cite{danilishinQuantumMeasurementTheory2012} 
\begin{align}
    \h I(t) &\propto \alpha^*(t) \h a(t) + \alpha(t) \h a^\dag(t) \\
    &= \sqrt2\abs{\alpha(t)}\h x_{\arg(\alpha(t))}(t).
\end{align}
where any noise from the local oscillator cancels out in the ideal case. This means that a phase-modulated local oscillator (with phase $\theta(t)=\arg(\alpha(t))$) can measure a chosen quadrature $\h I(t)\propto\h x_{\theta(t)}(t)$ at each time $t$. \\

By Eq.~16, the optimal compatible estimates $\h A$ and $\h B$ for a given frequency $\Om=\Delta$ are each a quadrature with time-dependent angle integrated against a kernel. While a given kernel could be realised using amplitude modulation (with amplitude $\abs{\alpha(t)}$), it can also simply be done in post-processing. Homodyne readout realises an individual measurement of either $\h A$ or $\h B$ at $\Om=\Delta$ in the single-parameter case of $w=0$ or 1. Since the quadrature angles of the integrands of $\h A$ and $\h B$ are different, however, a careful approach is required to simultaneously estimate $\h A$ and $\h B$. We cannot just measure $\h x_{\vartheta_A(t)}(t)$ and $\h x_{\vartheta_B(t)}(t)$ from Eq.~16 since they do not commute at every time $t$. Only the integrals against the kernels commute.

\subsection{Introducing compensating ancillae}

One method to simultaneously measure $\h A$ and $\h B$ using homodyne readout is to introduce ancillae to compensate for the nonzero commutator of $\h x_{\vartheta_A(t)}(t)$ and $\h x_{\vartheta_B(t)}(t)$ at each time $t$.\\

Let us consider the toy model since it is equivalent if we are focused on the signal at a given frequency $\Om=\Delta$. Suppose that to measure the incompatible displacements $\h X_1$ and $\mu \h P_1 + \sqrt{1-\mu^2} \h X_2$ we linearly introduce arbitrary Hermitian ancillae $\h Q_1$ and $\h Q_2$ and measure the following 
\begin{align}\label{eq:LinearAncilla}
    a \h X_1 + b \h Q_1, \quad c\left( \mu \h P_1 + \sqrt{1-\mu^2} \h X_2\right) + d \h Q_2
\end{align}
where $a$, $b$, $c$, and $d$ are nonzero real coefficients to be discovered. We introduced the ancillae to compensate for the commutator which implies if the ancillae commute with the system but not each other that
\begin{align}\label{eq:AncillaeCommutator}
    [\h Q_1, \h Q_2] = -i\mu \frac{a c}{b d}.
\end{align}
By calculating the signal-to-noise ratio for each measurement, the figure-of-merit for this measurement is
\begin{align}
    \Sigma' &=
    \frac{2w}{\left(\frac{a^2}{a^2\frac{1}{2} + b^2 \var{\h Q_1}}\right)} + \frac{2(1-w)}{\left(\frac{c^2}{c^2\frac{1}{2} + d^2 \var{\h Q_2}}\right)}
\end{align}
where we assume that the ancillae are statistically independent of the system. If $\Sigma'=\Sigma_H'$ to saturate the HCRB in Eq.~11, then 
\begin{align}
    \frac{a^2}{a^2 + b^2 2\var{\h Q_1}} &= \cos^2\left(\barBetter{\phi}\right)\\
    \frac{c^2}{c^2 + d^2 2\var{\h Q_2}} &= \cos^2\left(\barBetter{\phi} + \nu\right)
    .
\end{align}
Substituting these relations into Eq.~\ref{eq:AncillaeCommutator} provides the following constraint on the ancillae
\begin{align}
    \var{\h Q_1}\var{\h Q_2}=\frac{1}{4\mu^2}\tan^2\left(\bar\phi\right)\tan^2\left(\bar\phi + \nu\right)\abs{[\h Q_1, \h Q_2]}^2
    .
\end{align}
But, the Heisenberg Uncertainty Principle states that 
\begin{align}
    \var{\h Q_1}\var{\h Q_2} &\geq \frac{1}{4}\abs{\ev{[\h Q_1, \h Q_2]}}^2
\end{align}
such that we require (assuming that the commutator of the ancillae is a c-number)
\begin{align}
    \frac{1}{\mu^2}\tan^2\left(\barBetter\phi\right)\tan^2\left(\barBetter\phi + \nu\right) \geq 1.
\end{align}

Numerically, we find that this condition only holds (and is an equality) for $\mu=1$ and any $w$. It fails for $\mu<1$. This does not mean that any method using ancillae will fail for $\mu<1$, only that the method in Eq.~\ref{eq:LinearAncilla} cannot work.

\subsection{Realisation for $\mu=1$: an asymmetric beamsplitter}

\begin{figure}[t]
    \includegraphics[width=0.9\columnwidth]{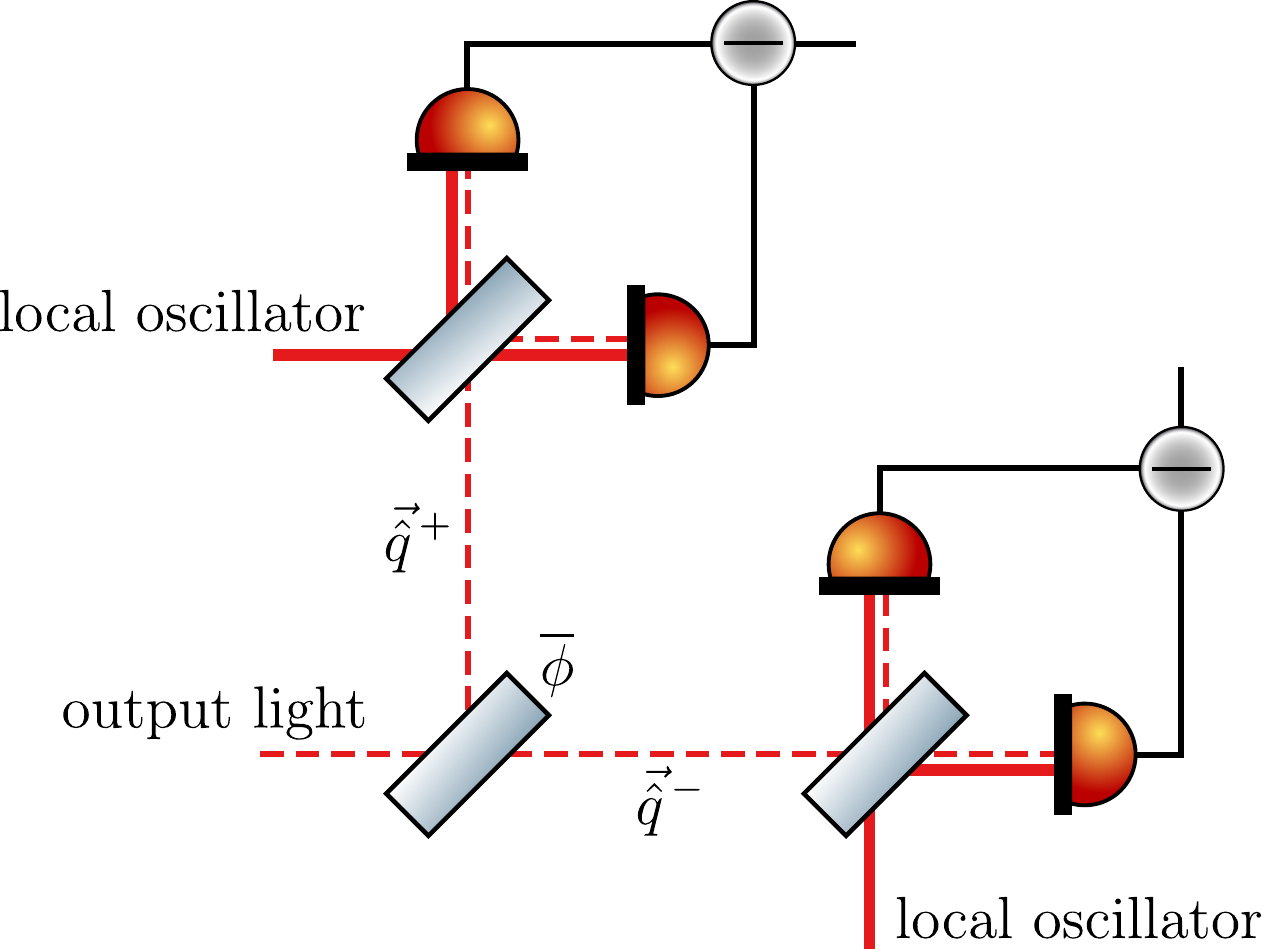}
    \caption{Measurement scheme for $\mu=1$ using an asymmetric beamsplitter and two homodyne readouts from Fig.~\ref{fig:S3}. The vacuum entering the other input port of the asymmetric beamsplitter is not shown.}
    \label{fig:S4}
\end{figure}

For $\mu=1$, the ancillae in Eq.~\ref{eq:LinearAncilla} correspond to mixing the output beam from the interferometer with vacuum using an asymmetric beamsplitter as shown in Fig.~\ref{fig:S4}. The parts for the two output beams from the beamsplitter are
\begin{align}
    \vec{\h q}^{\,+} &= \cos\left(\barBetter\phi\right) \vec{\h q} + \sin\left(\barBetter\phi\right) \vec{\h q}^{\,(0,\text{BS})} \\
    \vec{\h q}^{\,-} &= -\sin\left(\barBetter\phi\right) \vec{\h q}  + \cos\left(\barBetter\phi\right) \vec{\h q}^{\,(0,\text{BS})}
\end{align}
where the beamsplitter angle is chosen to be $\barBetter\phi$, $\vec{\h q}$ is the output beam from the interferometer in Eq.~10, and $\vec{\h q}^{\,(0,\text{BS})}$ is the vacuum entering the other input port of the beamsplitter. 
This realisation works only for $\mu=1$ because for $\mu=1$ the denominators in Eq.~11 are $\cos\left(\barBetter\phi\right)^2$ and $\sin\left(\barBetter\phi\right)^2$ which sum to unity. This is precisely the energy conservation required of a beamsplitter's power reflection and transmission coefficients $\abs{R}+\abs{T}=1$.
 
On the output beams from the beamsplitter $\vec{\h q}^{\,\pm}$, we can measure $\h A_\text{only}$ and $\h B_\text{only}$ in Eq.~6 using two homodyne readouts. The unbiased estimates for $A$ and $B$, respectively, are
\begin{align}
    \frac{\vec{d}_A\cdot\vec{\h q}^{\,+}}{\cos\left(\barBetter\phi\right)\eta^2}, \quad -\frac{\vec{d}_B\cdot\vec{\h q}^{\,-}}{\sin\left(\barBetter\phi\right)\eta^2}
\end{align}
which can be shown to saturate the HCRB and be compatible not just overall but at all times.

\subsection{General realisation using heterodyne readout}

In this section, we show that by measuring the output beam using homodyne readout with a phase-modulated local oscillator we can measure any two compatible linear combinations of $\vec{\h q}$. This includes our optimal estimates $\hat{A}$ and $\hat{B}$. This establishes a general measurement scheme that saturates the HCRB for any values of $\mu$ and $w$.\\

Suppose that we measure and record the timeseries of $\h x_{\vartheta_A(t)}(t)$ using homodyne readout with a phase-modulated local oscillator. The DC Fourier component of the timeseries times the kernel $c_A(t)$ is equal to $\h A$ by  Eq.~16. But, this is not all of the available spectral information. In particular, an arbitrary real linear combination with coefficients $c_{1}$ and $c_{2}$ 
of the real/imaginary parts, respectively, of the $2\Om$ Fourier component of the timeseries times $c_A(t)$ is, by Eq.~\ref{eq:OptMeasAsModulatedQuadsBetas},
\begin{align}
    \h q_\text{het} = 
    \intginf{t}&\left(c_{1} \cos(2\Om t) + c_{2} \sin(2\Om t)\right)
    \\&\cdot
    \left(
    \beta_{1,1}(t)\hat x(t) + \beta_{1,2}(t)\hat p(t)
    \right)   . \nonumber
\end{align}
Simplifying this into oscillations at the difference ($\Om$) and sum ($3\Om$) frequencies provides that
\begin{align}
    \label{eq:HetAsQ}
    \h q_\text{het}&=\sqrt{\pi T}\left(\vec{c}_\Om\cdot \vec{\hat q}(\Om)+\vec{c}_{3\Om}\cdot\vec{\hat q}(3\Om)\right)
    \where
    \vec{c}_{\Om}&= \frac{1}{2} \left(
    c_{1}
    \bmatrixByJames{
        \text{C}_{1,1} \\
        \text{C}_{1,2} \\ 
        -\text{C}_{1,3} \\ 
        -\text{C}_{1,4} \\
    }
    + c_{2}
    \bmatrixByJames{
        \text{C}_{1,3} \\ 
        \text{C}_{1,4} \\
        \text{C}_{1,1} \\
        \text{C}_{1,2} \\
    }
    \right)\label{eq:COmega}\\
    \vec{c}_{3\Om}&= \frac{1}{2}\left(
    c_{1}
    \bmatrixByJames{
        \text{C}_{1,1} \\
        \text{C}_{1,2} \\ 
        \text{C}_{1,3} \\ 
        \text{C}_{1,4} \\
    }
    + c_{2}
    \bmatrixByJames{
        -\text{C}_{1,3} \\ 
        -\text{C}_{1,4} \\
        \text{C}_{1,1} \\
        \text{C}_{1,2} \\
    }
    \right) 
\end{align}
where $\text{C}$ is from Eq.~\ref{eq:VectorEst}.\\

We want to show that measuring $\h q_\text{het}$ provides an estimate of $\h B$. The two vectors of $\vec{c}_\Om$ in Eq.~\ref{eq:COmega} commute with $\h A$ and are linearly independent of each other and $\h A$. Together with $\h A$, therefore, they span the space of all linear combinations of $\vec{\h q}$ that commute with $\h A$. Since $\h B$ commutes with $\h A$, there always exists real $\barBetter{c_{1}}$, $\barBetter{c_{2}}$, and $\barBetter{c_{3}}$ such that Eq.~\ref{eq:HetAsQ} becomes
\begin{align}\label{eq:HeterodyneNoise}
    \h q_\text{het} &= \left(\h B - \barBetter{c_{3}} \h A\right)+ \h B_\text{het}
    \where
    \h B - \barBetter{c_{3}} \h A &= \sqrt{\pi T}\vec{\barBetter{c_{\Om}}}\cdot\vec{\hat q}(\Om)\\
    \vec{\barBetter{c_{\Om}}}
    &=\vec{c}_{\Om}|_{c_{1}\mapsto\barBetter{c_{1}},c_{2}\mapsto\barBetter{c_{2}}}
    \intertext{and}
    \label{eq:HetNoiseFirst}
    \h B_\text{het} &= \sqrt{\pi T}\vec{\barBetter{c_{3\Om}}}\cdot\vec{\hat q}(3\Om) \\
    \vec{\barBetter{c_{3\Om}}}
    &=\vec{c}_{3\Om}|_{c_{1}\mapsto\barBetter{c_{1}},c_{2}\mapsto\barBetter{c_{2}}}.
\end{align}
Numerically, it appears that $\barBetter{c_{3}}=0$ here which is not necessarily true for two arbitrary commuting vectors. The non-stationary heterodyne noise $\h B_\text{het}$ at $3\Om$ commutes with and is statistically independent of $\vec{\hat q}(\Om)$ and therefore $\h A$ and $\h B$. Measuring $\h x_{\vartheta_A(t)}(t)$ and simultaneously estimating $\h A$ ($\h B$) from the DC ($2\Om$) Fourier component of the timeseries times $c_A(t)$ will not saturate the HCRB, unless the heterodyne noise can be squeezed without affecting $\h A$ and $\h B$. We will prove that this is possible.\\

For the detuned interferometer, we are interested in $\Om=\Delta$. Only saturating the HCRB at $\Om=\Delta$ is sufficient if we can assume that the signal is narrowband too. This is predicted for post-merger gravitational-wave signals at a given time~\cite{rezzolla2016gravitational}. E.g., if $\Om=\Delta=2\pi\times1000$~Hz, then although there can be 1~kHz and 3~kHz signals from the same source, they are emitted at different stages, separated in time. This motivates using a short-time Fourier transform in practice. We assume a monochromatic signal at $\Delta$ such that $\h B_\text{het}$ is at $3\Delta$ henceforth.

\subsection{Squeezing the heterodyne noise}

The time-domain expansion of $\h B_\text{het}$ in Eq.~\ref{eq:HeterodyneNoise} is similar to Eq.~16; it is equivalent to a kernel times some quadrature with a time-dependent angle oscillating at $3\Delta$. If this quadrature could be squeezed at all times, then $\h B_\text{het}$ would be squeezed. Although sufficient, however, this is not necessary, and a priori it could also affect $\h A$ and $\h B$. To determine whether $\h A$ and $\h B$ are affected, we would need to model the system. The Heisenberg-Langevin equations-of-motion with a time-dependent pump phase, however, are not simply solved in the Fourier domain because of convolutions arising with the pump. One solution method is to use a linear ansatz to turn the operator equations into differential equations such that a numerical solution can be found. Instead, we will use the one-photon formalism introduced previously to arrive at a solution using fixed pump phases.\\

Expanding Eq.~\ref{eq:HetNoiseFirst} into the one-photon formalism using Eq.~\ref{eq:OnePhotonFormalism} yields that
\begin{align}
    \h B_\text{het}
    &= \text{C}_\text{one}\bmatrixByJames{
        \h x'_{\om_0+3\Delta} \\
        \h p'_{\om_0+3\Delta} \\
        \h x'_{\om_0-3\Delta} \\
        \h p'_{\om_0-3\Delta} \\
    } \where
    \text{C}_\text{one}
    &= \frac{1}{4}
    \left(
    \barBetter{c_{1}}
    \bmatrixByJames{
    \text{C}_{1,1}- \text{C}_{1,4} \\
    \text{C}_{1,2}+ \text{C}_{1,3} \\
    \text{C}_{1,1}+ \text{C}_{1,4} \\
    \text{C}_{1,2}- \text{C}_{1,3}   
    }
    - \barBetter{c_{2}}
    \bmatrixByJames{
    \text{C}_{1,3}+ \text{C}_{1,2} \\
    \text{C}_{1,4}- \text{C}_{1,1} \\
    \text{C}_{1,3}- \text{C}_{1,2} \\
    \text{C}_{1,4}+ \text{C}_{1,1}   
    }    
    \right)    
    .
\end{align}
To squeeze $\h B_\text{het}$, therefore, the one-photon quadratures $\h x'_{\theta_\pm,\om_0\pm3\Delta}$ need to be each squeezed where the fixed quadrature angles are
\begin{align}\label{eq:NHeterodyneNoiseOnePhotonSqzAngles}
    \theta_\pm = \arctan\left(\frac{\barBetter{c_{1}} (\text{C}_{1,2}\pm \text{C}_{1,3})-\barBetter{c_{2}} (\text{C}_{1,4}\mp \text{C}_{1,1}) }{\barBetter{c_{1}} (\text{C}_{1,1}\mp \text{C}_{1,4})-\barBetter{c_{2}} (\text{C}_{1,3}\pm \text{C}_{1,2})}\right)
    .
\end{align}
Since the one-photon and two-photon formalisms coincide at $\Omega=0$ by Eq.~\ref{eq:OnePhotonFormalism}, squeezing a quadrature at $\Omega=0$ in one formalism also squeezes it in the other formalism. This means that performing degenerate (one-mode) squeezing at DC in the two-photon formalism centred at $\om_0+3\Delta$ with squeezing angle $\theta_+$ and then in the two-photon formalism centred at $\om_0-3\Delta$ with squeezing angle $\theta_-$ will suffice to squeeze $\h B_\text{het}$ as long as the two operations negligibly affect each other. Furthermore, $\h A$ and $\h B$, which are at $\om_0\pm\Delta$ in the one-photon formalism, should be negligibly affected if the two squeezing operations are sufficiently narrowband. The rest of this section will prove that this operation is possible using two squeezed, detuned, and narrowband filter cavities with different pump frequencies and phases as shown in Fig.~1e. Let the HWHM bandwidth of the filter cavities be $\gamma_\text{filt}\ll\Delta$ such that they are narrowband. We emphasise that this proposal of an experimental realisation demonstrates that a physical device can perform the optimal measurement in theory. We defer an exhaustive feasibility study to future work.

\subsection{Hamiltonian model of a squeezed cavity}
Suppose that a beam $\vec{\h q}^{(\omega)}$ described by the two-photon formalism centred at $\omega$ is reflected off a lossless cavity that is resonant at $\omega$ and $2\omega$. A crystal with a $\chi^{(2)}$ nonlinearity is placed within the cavity and pumped with a bright beam at $2\omega$ to form a degenerate (one-mode) optical parametric oscillator. The Hamiltonian for the interaction under the undepleted pump approximation is
\begin{align}
    \h H_\text{int} &=
    \frac{\hbar}{2}\chi\left(e^{i\theta}\left(\h a^{\text{(cav)}^\dag}\right)^2+e^{-i\theta}\left(\h a^\text{(cav)}\right)^2\right)
\end{align}
where $\chi$ is the squeezing parameter (proportional to the pump amplitude and interaction strength), $\theta$ is the pump phase, and $\h a^\text{(cav)}$ is the annihilation operator for the cavity mode. 
Assuming that the pump phase is fixed, then the Heisenberg-Langevin equations-of-motion can be solved linearly in the Fourier domain. The two-photon quadratures of the reflected beam, given those of the input beam, are 
\begin{align}\label{eq:SqzTunedCavIO}
    \bmatrixByJames{\h x^{(\omega, \text{out})}(\Om) \\ \h p^{(\omega, \text{out})}(\Om)} &= 
    \text{M}_\text{sqz}^\text{quad}(\Omega, \theta)
    \bmatrixByJames{\h x^{(\omega, \text{in})}(\Om) \\ \h p^{(\omega, \text{in})}(\Om)}
    \where
    \text{M}_\text{sqz}^\text{quad}(\Omega, \theta)&=
        \frac{1}{(\gamma_\text{filt} -i \Om )^2 - \chi ^2}\\&\cdot\bmatrixByJames{
     m_{1,1} & -2 \gamma_\text{filt}  \chi  \cos (\theta ) \\
     -2 \gamma_\text{filt}  \chi  \cos (\theta ) & m_{1,1} \\
    }\nonumber \\
    m_{1,1} &= \gamma_\text{filt} ^2+2 \gamma_\text{filt}  \chi  \sin (\theta )+\chi ^2+\Om ^2
    .
\intertext{Decomposing the quadratures into the real/imaginary parts provides a similar transformation}
\label{eq:PartsSqzTunedCavIO}
    \vec{\h q}^{(\om, \text{out})}(\Om)&=\text{M}_\text{sqz}(\Om, \theta)\vec{\h q}^{(\om, \text{in})}(\Om)
\end{align}
where $\text{M}_\text{sqz}$ is derived directly from $\text{M}_\text{sqz}^\text{quad}$ but we omit the full expression for brevity. $\text{M}_\text{sqz}$ does not explicitly depend on the two-photon centre frequency $\omega$, expect that the offset frequency $\Omega$ is defined with respect to $\omega$. 
From inspecting the covariance matrix, the quadrature at angle $\frac{\theta}{2}+\frac{\pi}{4}$ is squeezed and reaches zero variance at the lossless threshold of $\chi=\gamma_\text{filt}$ where gains balance losses. This means that if the cavities do not affect each other, then the squeezing angles $\theta_\pm$ in Eq.~\ref{eq:NHeterodyneNoiseOnePhotonSqzAngles} correspond to pump phases of
\begin{align}\label{eq:NHeterodyneNoiseOnePhotonPumpAngles}
    \theta_\pm^\text{pump} = 2\theta_\pm-\frac{\pi}{2}.
\end{align}

\subsection{Hamiltonian model of the filter cavities}

We derive the measured parts $\vec{\h q}^{(\om_0, \text{meas})}(\Om)$ after the two filter cavities in terms of the beam from the interferometer $\vec{\h q}^{(\om_0, \text{from IFO})}(\Om')$ (i.e.\ $\vec{\h q}$ in Eq.~4). The squeezing transformation in Eq.~\ref{eq:PartsSqzTunedCavIO} requires us to work in the two-photon formalism centred at the resonant frequency of each cavity. We need to use Eq.~\ref{eq:Shift2PhotonParts}, therefore, to shift from $\om_0$ to $\om_0+3\Delta$ (with $\delta\omega=3\Delta$), then to $\om_0-3\Delta$ (with $\delta\omega=-6\Delta$), and finally back to $\om_0$ (with $\delta\omega=3\Delta$). In the two-photon formalism centred at $\om_0$, the measured light at $\Om$ is a mixture of the following different frequencies $\Om'$ of the input light
\begin{align}\label{eq:MixedFrequencies}
    \Om'&=\Om,\, 6\Delta-\Om,\, 6\Delta+\Om,\, 12\Delta-\Om,\, 12\Delta+\Om
    \intertext{for example, at $\Om=\Delta$ for $\h A$ and $\h B$}
    \Om'&=\Delta,\, 5\Delta,\, 7\Delta,\, 11\Delta,\, 13\Delta
    \intertext{or, at $\Om=3\Delta$ for $\h B_\text{het}$}
    \Om'&=3\Delta,\, 9\Delta,\, 15\Delta
    .
\end{align}
Without the filter cavities or squeezing, any frequencies $\Om'\neq\Om$ cancel out and the transformation is proportional to the identity. With the filter cavities, however, the different frequencies do not cancel out from the final result. This is similar, but not identical, to how a time-dependent pump phase leads to convolutions in the Fourier transform of the two-photon formalism equations-of-motion. In the relevant narrowband regime, however, the frequency mixing is negligible. To summarise, the transformation is

\begin{align}
    \label{eq:OutputBeamWithoutQHorned}
    \vec{\h q}^{(\om_0, \text{meas})}(\Om)
    =
    \text{M}_\text{filt}(\Om,\theta_+^\text{pump},\theta_-^\text{pump})
    \vec{\h Q}^{(\om_0, \text{from IFO})}(\Om)
\end{align}
where the full expression for $\text{M}_\text{filt}$ is cumbersome and omitted for brevity but can be simply derived directly from $\text{M}_\text{shift}$ in Eq.~\ref{eq:Shift2PhotonParts} and $\text{M}_\text{sqz}$ in Eq.~\ref{eq:PartsSqzTunedCavIO}.
$\vec{\h Q}^{(\om_0, \text{from IFO})}(\Om)$ is a 20-component vector comprising the five $\vec{\h q}^{(\om_0, \text{from IFO})}(\Om')$ for $\Om'$ in Eq.~\ref{eq:MixedFrequencies}.\\

The pump phases from Eq.~\ref{eq:NHeterodyneNoiseOnePhotonPumpAngles} used in Eq.~\ref{eq:OutputBeamWithoutQHorned} assume that the two cavities do not affect one another. The detuning of the second cavity, however, slightly rotates the squeezed light from the first cavity. This means that the first pump phase $\theta_+$ needs to be counter-rotated to compensate. Although the second cavity is narrowband, the highly squeezed state is sensitive to even small misalignments of the quadrature angle. We show this numerically by optimising for the measured variance of $\h B_\text{het}$ from $\vec{\h q}^{(\om_0, \text{meas})}(\Om)$. This yields that $\theta_+$ in Eq.~\ref{eq:NHeterodyneNoiseOnePhotonPumpAngles} is misaligned from the optimal angle by $\mathcal{O}(0.01)$ radians but that $\theta_-$ is optimal. With the optimal pump phases, the variance of $\h B_\text{het}$ can be reduced to zero at threshold in the lossless model. We defer studying the effect of losses to future work.\\

With the heterodyne noise squeezed, our numerical model also shows that $\h A$ and $\h B$ are minimally affected for sufficiently narrowband cavities. E.g., for $\gamma_\text{filt}=2\pi\times30$~Hz and $\Delta=2\pi\times3000$~Hz, the variances of $\h A$ and $\h B$ are unaffected, their expectation values change by roughly one part in 100, and there is mixing in of roughly one part in 10000 of the $5\Delta$ and $7\Delta$ components in Eq.~\ref{eq:MixedFrequencies}. Given any tolerance, there exist (in theory) sufficiently narrowband cavities such that $\h A$ and $\h B$ are unaffected. \\

The narrowband cavities present some experimental challenges. To achieve the narrowband limit of saturating the HCRB, for each filter cavity, the power transmission of the input mirror must be low and/or the length of the cavity must be long. E.g, for $\gamma_\text{filt}=2\pi\times30$~Hz, a transmission of 1\% requires 4-km long filter cavities, comparable to the arms of the interferometer. Unlike the arm cavities, however, these filter cavities can and should be folded such that no gravitational-wave signal is accrued within them. Together with even lower transmission, this should allow for the filter cavities to be achieved at the table-top scale. This means that the HCRB can be saturated by this scheme.\\

\emph{Necessity of squeezing} --- Is there additional information about $A$ or $B$ for a given frequency $\Om=\Delta$ in the spectrum of the recorded timeseries of $\h x_{\vartheta_A(t)}(t)$ at other frequencies? If true, then is there a way that this could be used to realise the optimal measurement without using squeezing? A priori, the squeezed states used above could be the only optimal measurement basis that is experimentally accessible, but we have not shown this. For the timeseries of $\h x_{\vartheta_A(t)}(t)$ times $c_A(t)$, then the DC and $2\Om$ components are accounted for in Eq.~\ref{eq:HetAsQ}. The $\Om'$ component of the output light contains $\vec{\h q}(\Om'\pm\Om)$. Since we want to estimate $\vec{\h q}(\Om)$, then $\Om'\pm\Om=\Om$ implies that only $\Om'=0$ or $2\Om$. There is no additional information, therefore, in the spectrum of $\h x_{\vartheta_A(t)}(t)$ times $c_A(t)$. The multiplication by $c_A(t)$, however, was only done in post-processing. By the Convolution Theorem, therefore, we have only used part of the available Fourier transform of $\h x_{\vartheta_A(t)}(t)$.

We defer to future work the calculation of the Fisher information with respect to $A$ and $B$ of measuring the continuum $\h x_{\vartheta_A(t)}(t)$ for all $t$ to definitively resolve these questions.

\subsection{Summary of the measurement protocol}
We summarise the combined strategy that we propose to experimentally saturate the HCRB for all $\mu$ and $w$:
\begin{outline}
    \1 For $\mu=0$, the interferometer is tuned and fixed-angle homodyne readout works at all $\Omega$.
    \1 For $w=\frac{1}{2}$, the standard variational readout scheme works at all $\Omega$.
    \1 For $w=0$ or 1 (single-parameter estimation), the homodyne measurement with a phase-modulated local oscillator works at $\Omega=\Delta$. 
    \1 For $\mu=1$, an asymmetric beamsplitter with homodyne measurements with phase-modulated local oscillators on each output beam works at $\Omega=\Delta$.
    
    \1 For $0<\mu\leq1$ and $0<w<1$, squeezing the output beam using two detuned and narrowband filter cavities before performing a homodyne measurement with a phase-modulated local oscillator works at $\Omega=\Delta$. The first signal phase is estimated from the DC Fourier component of the timeseries multiplied by a kernel. The second phase is estimated from the $2\Delta$ Fourier component and requires squeezing to eliminate the heterodyne noise at $3\Delta$.
    
\end{outline}

\section{Alternative approaches to find an experimental realisation}
This section presents other attempts that we have made to realise the optimal measurement scheme. We discuss these directions to explain some of the complexity of the proposal presented above and in the Letter.

\subsection{Simplifying the measurement scheme} 
Another idea is that, since $\h A$ and $\h B$ commute, a symplectic 4-by-4 matrix $\text{M}_\text{simple}$ exists to simplify the coefficients in Eq.~\ref{eq:VectorEst} to\begin{align}\label{eq:SymplecticToDecomp}
    \text{M}_\text{simple} \text{C}_{1,.}^\text{T} \propto \bmatrixByJames{1 \\ 0 \\ 0 \\ 0}, \quad 
    \text{M}_\text{simple} \text{C}_{2,.}^\text{T} \propto \bmatrixByJames{0 \\ 0 \\ 1 \\ 0}
\end{align}
where $\text{C}_{j,.}$ is the $j$th row of $\text{C}$ and the vectors on the right-hand side are with respect to $\vec{\h q}$. Under this transformation, the estimates would become
\begin{align}
    \h A \propto \re{\h x(\Delta)}, \quad 
    \h B \propto \im{\h x(\Delta)}
\end{align}
such that they can both be inferred from a measurement of $\h x(t)$ for all $t$.\\

This transformation does exist. The biased optimal measurements for the toy model in Eq.~13 can be transformed symplectically as
\begin{align}
    \text{M}_\text{simple}' &= \bmatrixByJames{
         \cos \left(\barBetter{\phi} \right) & 0 & 0 & -\sin \left(\barBetter{\phi} \right) \\
         0 & \cos \left(\barBetter{\phi} \right) & \sin \left(\barBetter{\phi} \right) & 0 \\
         0 & -\sin \left(\barBetter{\phi} \right) & \cos \left(\barBetter{\phi} \right) & 0 \\
         \sin \left(\barBetter{\phi} \right) & 0 & 0 & \cos \left(\barBetter{\phi} \right) \\
    }
    \intertext{such that}
    \text{M}_\text{simple}'\;\h T_1 &= \bmatrixByJames{1\\0\\0\\0\\}, \quad
    \text{M}_\text{simple}'\;\h T_2 = \bmatrixByJames{0\\0\\1\\0\\}
\end{align}
where the vectors on the right-hand side are with respect to $\vec{\h X}$.
Recall that $\hat{A}$ and $\hat{B}$ in terms of the toy model estimates are
\begin{align}\label{eq:OptEstWrtT}
    \hat A = \frac{1}{\eta\cos\left(\barBetter{\phi}\right)}\hat T_1,\quad
    \hat B = \frac{1}{\eta\cos\left(\barBetter{\phi}+\arcsin(\mu)\right)}\hat T_2.
\end{align}
The symplectic transformation in Eq.~\ref{eq:SymplecticToDecomp}, therefore, is 
\begin{align}
    \text{M}_\text{simple}=\text{M}_\text{simple}'\text{M}
\end{align}
where $\text{M}$ from Eq.~\ref{eq:MAtDetFreq} is used to transform the estimates in Eq.~\ref{eq:OptEstWrtT} from $\vec{\h q}$ to $\vec{\h X}$ by Eq.~10.\\

Numerically calculating $\text{M}_\text{simple}$ for the parameters in the Letter shows that it is a double-rotation of $\R^4$ with eigenvalues $e^{\pm i \theta_1}$ and $e^{\pm i \theta_2}$ for some $\theta_1\neq\theta_2$. In block-diagonal form, it is
\begin{align}\label{eq:BlockDiagonalForm}
    \text{M}_\text{simple}= \text{S}^{-1}
    \bmatrixByJames{
        \text{Rot}(\theta_1) & 0 \\
        0 & \text{Rot}(\theta_2)
    }   
    \text{S}
\end{align}
where $\text{Rot}$ is the 2-by-2 rotation matrix and the similarity transformation $\text{S}$ does not have a complex compact form, i.e.\ there does not exist some complex 2-by-2 matrix $\text{S}_\C$ such that
\begin{align}
\text{S} = \bmatrixByJames{
    \re{S_\C} & \im{S_\C} \\
    -\im{S_\C} & \re{S_\C}
}.
\end{align}

The problem is that it is not clear how to physically realise $\text{M}_\text{simple}$. We want a decomposition into operations that we know how to perform, e.g.\ phase delays and beamsplitters. A phase delay is the symmetric case of Eq.~\ref{eq:BlockDiagonalForm}
\begin{align}
    \bmatrixByJames{
        \text{Rot}(\theta) & 0 \\
        0 & \text{Rot}(\theta)
    }.
\end{align}
But, we do not know how to experimentally realise the asymmetric case. This is a direction that future work could continue with.

\subsection{Using integrating ancillae}
One alternative method is to consider two ``integrating'' ancillae whose evolution naturally performs the integration in Eq.~16 and which can be measured at $t=T$ to obtain $\h A$ and $\h B$ directly. E.g., consider reflecting the output beam off two tuned and squeezed cavities. Within each cavity, a $\chi^{(2)}$ crystal with a time-dependent pump creates nondegenerate (two-mode) squeezed states. Each parametric interaction entangles the ``signal'' mode (the output beam from the interferometer) with an ``idler'' mode (a spatially separable optical mode, e.g.\ at a different frequency). The intracavity idler modes are integrating ancillae since the cavity naturally integrates the light arriving at different times. At $t=T$, the idler mode leaking out from each cavity can be measured. Whether this can realise the optimal measurement depends on if the initial conditions and vacuum modes driving the idler modes are negligible at $t=T$. We defer investigating approach this to future work.